%% file: main.tex
\documentclass[acmsmall,screen]{acmart}

\AtBeginDocument{%
  \providecommand\BibTeX{{%
    \normalfont B\kern-0.5em{\scshape i\kern-0.25em b}\kern-0.8em\TeX}}}

\copyrightyear{2020}

\setcopyright{acmcopyright}
\acmJournal{PACMHCI}
\acmYear{2021} \acmVolume{5} \acmNumber{CSCW1} \acmArticle{90}
\acmMonth{4} \acmPrice{15.00}\acmDOI{10.1145/3449164}

\usepackage{xspace}
\usepackage{subcaption}
\usepackage{graphicx} 
\usepackage{multirow}
\usepackage{booktabs}
\usepackage{array}
\usepackage{datetime}

\received{June 2020}
\received[revised]{October 2020}
\received[accepted]{December 2020}

\usepackage{appendix}
\begin{document}

\title{Analyzing Twitter Users' Behavior Before and After Contact by Russia's Internet Research Agency}



\author{Upasana Dutta, Rhett Hanscom, Jason Shuo Zhang, Richard Han, Tamara Lehman, Qin Lv, Shivakant Mishra}
\affiliation{
  \institution{ University of Colorado Boulder}
  \city{Boulder}
  \state{Colorado}
  \postcode{80309}
  \country{USA}
}
\email{Upasana.Dutta@colorado.edu, Rhett.Hanscom@colorado.edu}

\renewcommand{\shortauthors}{Dutta and Hanscom et al.}


\input{abstract}



\newcommand{\para}[1]{\noindent{\bf #1}}
\newcommand{\figref}[1]{Figure \ref{#1}}
\newcommand{\equationref}[1]{Equation \ref{#1}}
\newcommand{\secref}[1]{Section \ref{#1}}
\newcommand{\tableref}[1]{Table \ref{#1}}

\graphicspath{{graphics/}}
\maketitle

\input{intro}

\input{background}

\input{data}

\input{method}

\input{results}

\input{conclusion}

\input{acknowledgments}


\bibliographystyle{ACM-Reference-Format}
\bibliography{bibliography}



\end{document}

%% file: abstract.tex


\begin{abstract}

Social media platforms have been exploited to conduct election interference in recent years.  In particular, the Russian-backed Internet Research Agency (IRA) has been identified as a key source of misinformation spread on Twitter prior to the 2016 U.S. presidential election. The goal of this research is to understand whether general Twitter users changed their behavior in the year following first contact from an IRA account. We compare the before and after behavior of contacted users to determine whether there were differences in their mean tweet count, the sentiment of their tweets, and the frequency and sentiment of tweets mentioning @realDonaldTrump or @HillaryClinton.  Our results indicate that users overall exhibited statistically significant changes in behavior across most of these metrics, and that those users that engaged with the IRA generally showed greater changes in behavior.

\end{abstract}

%% file: intro.tex

\section{Introduction}

For many years, the growth of online social media has been seen as a tool used to bring people together across borders and time-zones.  However, malicious actors have been shown to have harnessed the tool of social media to instead test their hand at sowing regional discord \cite{washingtonpost,woolley2017computational}.  Prior to the 2016 U.S.  presidential election, the Internet Research Agency (IRA), a Russian organization, created a “troll farm”, which registered and deployed thousands of sham accounts across various social media platforms \cite{us2018exposing}.  These IRA accounts have been shown to have both targeted U.S. citizens and worked towards goals of forging interference in American politics, specifically the 2016 U.S. presidential election involving primary candidates Hillary Clinton and Donald Trump \cite{us2018exposing}. Their operations were deeply sophisticated and mainly aimed at instigating the social media users against each other on socio-economic grounds, political sentiments, and voting perceptions \cite{wiredpost}. 

The IRA deployed 3,841 fake accounts on Twitter, a micro-blogging site, which resulted in 1.4 million Twitter users being exposed to the IRA tweets, as reported by Twitter \cite{twitterpost}, which led to roughly 73 million total engagements \cite{engagementpost}. The main goal of this research is to understand to what extent the behavior of Twitter users changed after contact by IRA bots during or prior to the 2016 U.S. presidential election cycle.  The implications are clearly enormous as a new U.S. presidential election of 2020 looms, which may be disrupted by the continuing efforts of the IRA \cite{ghanapost} or other like-minded organizations.  This work focuses on an extensive before and after analysis of Twitter users' behavior based on six quantifiable metrics to identify the extent of changes following initial IRA contact.

While prior work has studied in great detail the behavior of the IRA bot accounts and techniques that they employed to spread misinformation~\cite{howardira,boatwright2018troll,boyd2018characterizing,keller2020political}, based on a data set released by Twitter~\cite{twitterpost}, much less work has examined the \emph{targets} of the IRA bots, namely the overall Twitter user population.  One recent work ~\cite{bail2020assessing} examining such targets surveyed the attitudes and behaviors of 1,239 Republican and Democratic Twitter users and found no evidence that Russian agents successfully changed user opinions or voting decisions. However, a limitation of this study is that it focused on a subset of Twitter users who already held fairly strong partisan beliefs and as a result may not be easily swayed by IRA bots.

Our interest is to study a broader population consisting of all the Twitter users contacted by the IRA to understand if their behavior changed before and after contact. This broader population of users may be less partisan in nature and hence more amenable to influence.  
Specifically, this study focuses on the following research questions:

\begin{itemize}
    \item \textbf{RQ1}: Did the contacted Twitter users exhibit a change in behavior following contact with IRA accounts?
    \item \textbf{RQ2}: Did the  contacted Twitter users who engaged back with the IRA accounts behave differently compared with those who did not after the contact?
    \item \textbf{RQ3}: Did the behavior of users with high follower counts differ from those with lower follower count after initial contact with the IRA?  
\end{itemize}

To answer these questions, we began with a data set released by Twitter, in which they identified 3,479 accounts as IRA bots \cite{twitterpost}.  Using these accounts, we then were able to find the users whom the IRA bots contacted through retweets, replies, and mentions. and then collected tweets of those Twitter users.  To identify whether there were significant changes in the tweeting behavior before and after their interaction with the IRA, and if so by how much, we selected objective behavioral metrics that measured monthly tweet count, mean sentiment of tweets, frequency of mentioning either @realDonaldTrump or @HillaryClinton, and the mean sentiment of tweets that tag either @realDonaldTrump or @Hillary Clinton.

In addition, our research considers whether the behavior of those contacted Twitter users with a high follower count ($\ge{5000}$) differed from the behavior of contacted users with a more typical follower count (< 5000).  Combined with our interest in studying users who responded to and did not respond to the contact attempts by the IRA, this creates four groups of users that we studied: responsive high follower count users, non-responsive high follower count users, responsive low follower count users, and non-responsive low follower count users.

\begin{figure}[t]
\centerline{\includegraphics[width=0.85\linewidth]{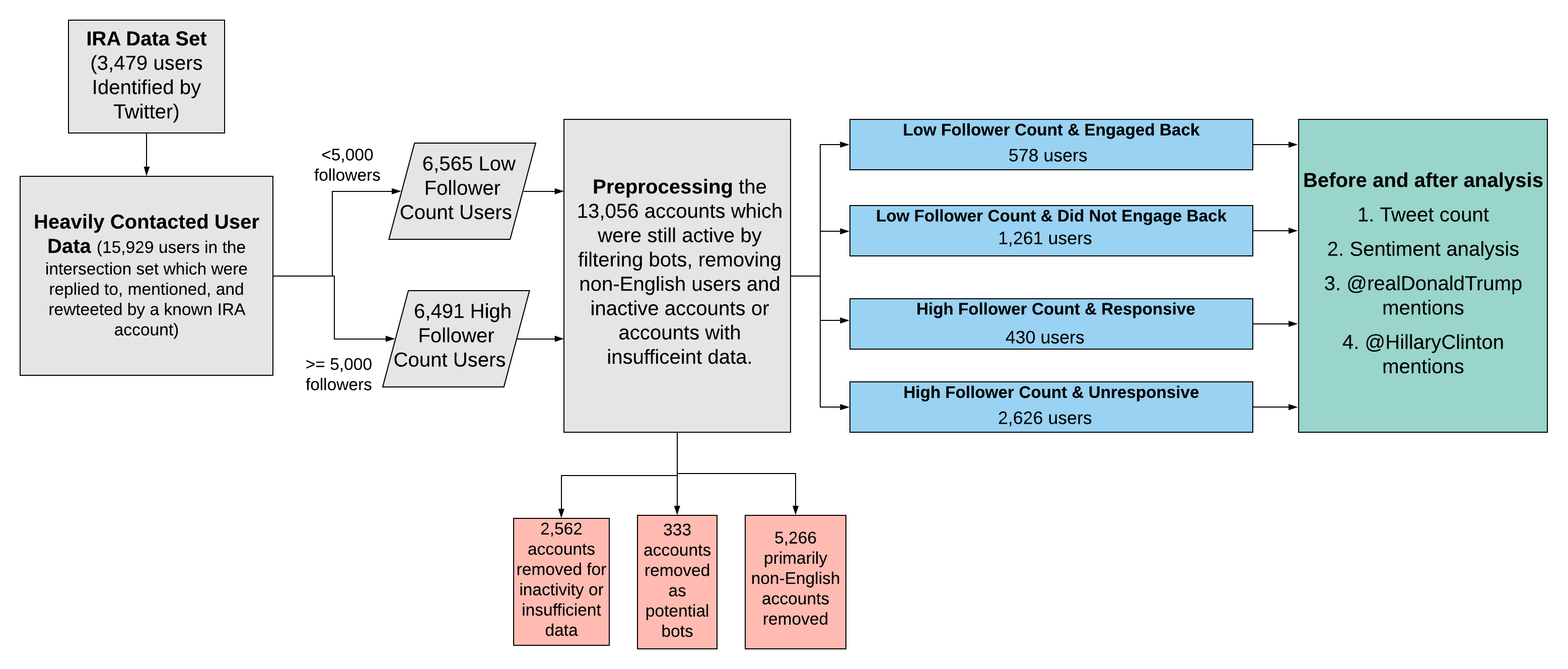}}
\caption{ The process we used to create and analyze the Twitter data set. The final user population has been divided into four subsets: responsive and unresponsive low follower count users, as well as responsive and unresponsive high follow count users. All users in this work were filtered to remove those with bot-like behavior, protected accounts, predominantly non-English tweets, or insufficient data to be studied. Before and after analysis was performed to evaluate monthly tweet volumes, overall sentiment, as well as the frequencies and sentiment of tweets mentioning either @realDonaldTrump or @HillaryClinton. } 
\label{filter_process_new}
\end{figure}

We further desired to restrict the data set to American Twitter users due to our interest in the 2016 U.S. `presidential election, but lacked geo-location or other information to determine whether a Twitter user is American.  Our closest approximation was to filter the data set for English language tweets, as the IRA bots also conversed with users using other languages, especially Russian.  In addition, the user pool was filtered in order to remove potential bots, protected accounts, and accounts with insufficient data to be studied. We also focused our efforts on studying those users who were most contacted by the IRA, namely those who had been contacted in all three ways through retweets, replies, and mentions by the IRA. We reasoned that if this group could first be shown to exhibit a change in behavior, then it would justify expanding our data collection to the much larger set of those users who had been contacted in at least one of the three ways.
\figref{filter_process_new} illustrates the overall process of data pre-processing and the before and after analysis.

Additionally, in order to identify any behavioral trends on the Twitter platform that may have effected the users contacted by the IRA, a baseline user set of 8,023 random Twitter users was compiled. Each of these users underwent the same filtering process that a user contacted by the IRA underwent (i.e., language analysis, Botometer check, etc.)

With these caveats in mind, this work provides some of the first evidence that contacted Twitter users' behavior underwent significant changes in a variety of ways following interactions with Russia's Internet Research Agency prior to the 2016 U.S. presidential election. Specifically, this paper makes the following contributions:

\begin{itemize}
    \item A substantially broader population of Twitter users is studied to understand whether user behavior changed before and after contact by the IRA accounts.
    \item Evidence is found of statistically significant increases in tweeting frequency across all user groups, including responsive and non-responsive high and low follower count users, after initial contact by the IRA. (RQ1)
    \item Three of the four user groups exhibited statistically significant decreases in sentiment (i.e., less positive or more negative) after initial contact by the IRA. (RQ1)
    \item The frequency of monthly mentions of @realDonaldTrump is shown to significantly increase for all user groups after initial contact by the IRA. (RQ1)
    \item Statistically significant increases in monthly mentions of @HillaryClinton occurred for all four user groups after initial contact by the IRA. (RQ1)
    \item Across a variety of different metrics, responsive users who engaged with the IRA generally showed change that was either higher in percentage or satisfied stricter statistical significance  compared with non-responsive users.  (RQ2)
    \item We found no statistically significant difference in the behavior of high versus low follower count users. (RQ3)
    \item Users contacted by the IRA we found to have significantly different behavior than that of the random baseline set in all areas except @realDonaldTrump mention sentiment of unresponsive high follower count users.
    
\end{itemize}

The intent of this paper is to identify and quantify correlation between initial contact by the IRA and changed behavior by contacted users where it exists.  The intent of the paper is not to establish causality, i.e., that IRA contact caused the changes in behavior, which would require a different path of research.  The Discussion section of the paper describes this topic in more detail.

In the following, we first describe background and related work, then explain our data collection and pre-processing filters, followed by a description of our method for performing before and after analysis, including the metrics used.  We then present the key results of our data analysis.  The paper finishes with a discussion of the implications and limitations of our work as well as a summary of our conclusions.

%% file: background.tex

\section{Background and Related Work}

\subsection{Internet Research Agency}
Internet Research Agency (IRA) is a well-known Russian company engaged in online influence operations on
behalf of Russian business and political interests \cite{kriel2019reverse, ruck2019internet, boatwright2018troll, diresta2018tactics}. 
According to the Mueller Report released in 2017 \cite{mueller2019report}, 
the IRA was using social media to undermine the electoral process and is at the center
of the federal indictment.
They have been collecting sensitive information, such as names, addresses, phone numbers, email addresses,
and other valuable data about thousands of Americans.
The IRA campaign during the 2016 U.S. presidential election has drawn great interest in the research community.
Earlier research has characterized the strategies of IRA activity on social media from various perspectives
\cite{howardira,boatwright2018troll,boyd2018characterizing,keller2020political}.
To the best of our knowledge, most of this work does not examine how IRA accounts interact with general
Twitter users, and whether or not these campaigns play a role in changing attitudes and behaviors of such users. 
\citet{bail2020assessing} used survey data that described the attitudes and behaviors of 1,239 Republican 
and Democratic Twitter users and found no evidence that these Russian agents 
successfully changed user opinions or voting decisions. 
In comparison, we have expanded the scope of our study to consider a much broader population of users over a longer observational period, and found significant 
differences in their behaviors before and after the contact.

\subsection{Malicious Bots on Social Media}
While some bots are designed for providing better service/management, such as auto-moderators and chat bots, 
they can be misused for bad behavior by extreme groups \cite{bessi2016social}. 
For instance, researchers have detected bots that spread misinformation \cite{shao2017spread,ferrara2016rise}, 
interfere with elections \cite{ferrara2017disinformation}, and amplify hate against minority groups \cite{albadi2019hateful}.
ISIS propagandists are also known for using bots to inflate their influence and 
promote extreme ideologies \cite{benigni2017online}. 
Due to bots' sophisticated behavior, it is increasingly important to understand them and investigate 
strategies to combat malicious bots in an online environment.

\subsection{Social Media Manipulation during Elections}
There is growing interest in understanding manipulation campaigns deployed on social media sites 
specifically targeting elections. 
Other than the IRA campaign that we have mentioned previously, 
similar approaches have been replicated by different countries. 
For example, during the 2016 UK Brexit referendum it was found that political bots played
a small but strategic role in shaping Twitter conversations \cite{howard2016bots}.
\citet{ferrara2017disinformation} studied the MacronLeaks disinformation campaign during
the 2017 French presidential election and found that most users who engaged in this campaign
were foreigners with preexisting interest in alt-right topics and alternative news media. 
\citet{arnaudo2017computational} examined computational propaganda during
three election related political events in Brazil from 2014 to 2016, and found that social media manipulation was becoming more diverse and at a much larger scale. 
Starbird et al. conducted an interpretative analysis 
of a cross-platform campaign targeting a rescue group in Syria,
which conceptualized what a
disinformation campaign is and how it works \cite{starbird2019disinformation,wilson2020cross}.

%% file: data.tex
\section{Data Collection}

The goal of this research is to understand behavioral changes of general American Twitter users after interactions with the IRA. To achieve this goal, IRA tweets were analyzed to find a sample of user population the IRAs interacted with on Twitter. To keep the scope of this work focused on the most heavily contacted users, only those users who were mentioned, replied to, as well as retweeted by the IRAs were considered for the analysis. The reason for this selection is that these users are potentially the ones most impacted by IRA interactions. To isolate American users who were contacted, the pool of potential users was put through a number of vetting processes in order to weed out users who did not have majority of their tweets published in English. The screening also removed users with protected accounts, users who had a high probability of being a Twitter bot account, and  users whose first accessible tweet occurred after IRA contact. This process has been illustrated in Figure ~\ref{filter_process_new}.


\textbf{IRA Data Set:} This study used an unhashed version of the data set made public by Twitter for research purposes \cite{twitterpost}. The IRA accounts were confirmed by Twitter to be Russian agents by the U.S. government during the 2017 investigation into the 2016 U.S. presidential election cycle \cite{us2018exposing}. The data set contains the tweets of 3,479 IRA accounts, that are assumed to be connected to the propaganda efforts of the Internet Research Agency and are currently suspended \cite{twitterpost}.

\textbf{Contacted User Data Set:} Twitter has identified 1.4 million users who interacted with the IRA via Twitter \cite{twitterpost}. However, in order to preserve the anonymity of these users, that data set has not been released to the public. Therefore, to study the users who interacted with the IRA, our research first analyzed a total of 8,768,633 tweets shared by the 3,479 IRA accounts, from May 2009 to June 2018, and identified users who were either replied to, retweeted by, or mentioned by an IRA account.  ~6\% of the IRA tweets were in reply to other users, resulting in 82,806 distinct users contacted by the IRAs.  Out of these users, 1,533 (~2\%) were identified as themselves being affiliated with the IRA.  ~38\% of the IRA tweets were retweets of other Twitter users, resulting in 204,273 distinct users who were retweeted by IRA accounts, out of which 1,819 (~1\% of them) were identified as themselves being affiliated with the IRA.  ~46\% of the IRA tweets mentioned other users, resulting in 829,080 distinct users who were mentioned by the IRA accounts. Among these users, 2,141 (~0.3\%) were identified as themselves being affiliated with the IRA.


In aggregate, we found 837,688 non-IRA Twitter users who received either  mentions, replies, or retweets from an IRA account. The set of highly contacted users that were contacted by all three methods (mention, reply, retweet) consists of 15,929 users, and is the focus of this study.  We plan to extend this study in the future to consider the wider set of 837,688 users contacted by at least one of the three means.


\begin{figure}[t]
\centerline{\includegraphics[width=0.85\linewidth]{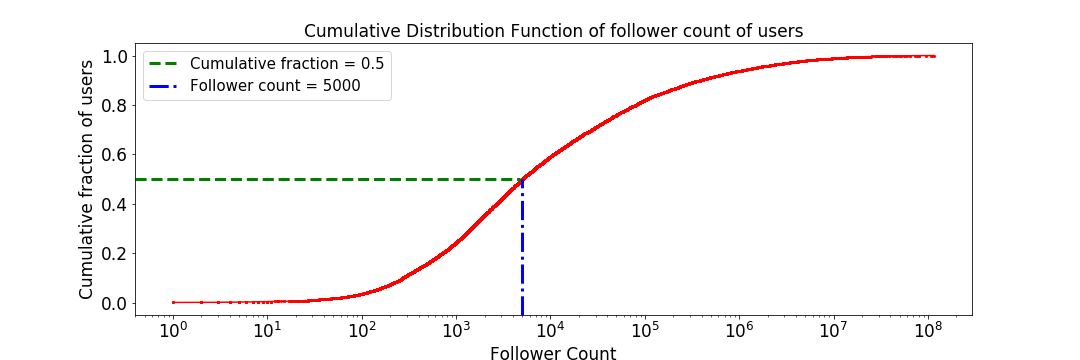}}
\caption{Cumulative Distribution Function of follower count of users. The x-axis shows the follower count in log-scale, the y-axis shows the fraction of 13,056 users who have a follower count less than or equal to the corresponding x-value. It can be inferred from the plot that approximately 50\% of the users have follower count $\leq 5000$.} 
\label{follower_distribution}
\end{figure}

\textbf{High vs. Low Follower Count Users:} We separated the set of users identified above into two groups, high and low follower count users, to better understand the IRA impact on users with varying following on Twitter.  As a follower count of 5,000 divided our user pool fairly well into two halves (\figref{follower_distribution}), 5,000 was chosen to be the threshold between high and low follower count groups. We consider high follower count users to be those users who have at least 5,000 followers. Past research pertaining to reach of users at various follower-levels (e.g., non-influencers, micro-influencers, and influencers) has also used this critical point of 5,000 followers to separate classes of influence \cite{lou2019influencer, kolo2018social}.

Tweepy, a Python library for the Twitter API \cite{tweepy}, was used to calculate each user's follower count. Tweepy was unable to access information on 2,873 protected, deleted, or similarly inaccessible Twitter users who were then discarded.  We found that of the remaining 13,056 accounts, there were 6,565 low follower count users and 6,491 high follower count users.

\textbf{Twitter User Data Scrape:} For these 13,056 users we utilized Twint, a Twitter-scraping Python library, to collect their user data~\cite{twint}.  Twint has less severe rate restrictions than Twitter and specializes in excavating archival data specifically~\cite{twint}.  Twint can compile all tweets publicly authored by a user but it is restricted in the retweet and favorite tweets that it can scrape. Twint can only retrieve the most recent retweets and the most recently favorite tweets~\cite{twint}. When working with data from previous years, it is unlikely that a majority of favorite tweets from said period will be reachable. For this reason, retweets and favorite tweets were not utilized within this analysis, as their presence could be unevenly distributed due to this time-sensitive constraint.

Twint is also limited in the state of the accounts it is able to retrieve.  Twint is unable to retrieve account data for deleted accounts \cite{twint}.  Additionally, Twint is unable to scrape tweets published while a Twitter account was ``protected'' (private), even if the account is later made public \cite{twint}.  Users ``shadow-banned'' from Twitter were also inaccessible via Twint \cite{twint}; ``Shadow-banning'' is a process wherein a user's data is not available through the Twitter search feature for a period of time \cite{twittershadow, nytshadow}.

\textbf{Responsive vs. Unresponsive Users:} After removing the deleted and inaccessible accounts within the low follower count user set, 5,001 users' tweet histories were scraped. Analysis of these tweets revealed that 2,285 of the users in this set (45.6\%) engaged back with the IRA by either mentioning an IRA account in their tweets or by replying to the tweets shared by an IRA agent, while the remaining 54.4\% of users showed no signs of engagement through mentions with an IRA account.  Within the high follower count user set, 5,712 users' tweet histories could be obtained.  Analysis of the tweets of this high follower count user set revealed that 1,054 of them (18.5\%) engaged back with IRA accounts by either mentioning the IRAs in their tweets or by replying to the tweets shared by the IRA, while the remaining 81.5\% of users showed no signs of engaging back with an IRA account.

\textbf{Removal of Potential Bots:} In order to ensure the users under the purview of this study were legitimate users, a number of filters were utilized to remove accounts suspected of deviating from this criteria. A Python API, Botometer, was used to assess the likelihood that any individual user was an automated account and not a natural user \cite{botometer}. Botometer uses 20 features from a user's profile along with supervised machine learning to identify accounts primarily run with the help of automation software. Among the greatest limitations to Botometer are botnets, or a number of social bots working together to share the same media \cite{doi:10.1002/hbe2.115}. Botnets use their coordinated posts to give the appearance of popularity to their messaging. Botnets have proven difficult to detect for both supervised machine learning algorithms and humans \cite{Cresci_2017}. A ``universal'' and an ``English'' score are returned by Botometer for each user, representing the probability that an account is run by a bot \cite{botometer}. The ``English'' score checks only English-specific information, while the ``universal'' score is not language specific \cite{botometer}.  Accounts that received probabilities of 0.4 or greater from Botometer were deemed suspected bots and thus removed from this analysis. A similar threshold was previously tested and shown to be a near-optimal critical threshold with Botometer, offering almost 80\% accuracy in bot detection \cite{wojcik2018bots}. 151 of the 5,001 users within this low follower count user set were removed by the Botometer low follower count, whereas 182 of the 5,712 users were removed from the high follower count user set.

\begin{figure}[t]
\centerline{\includegraphics[width=0.85\linewidth]{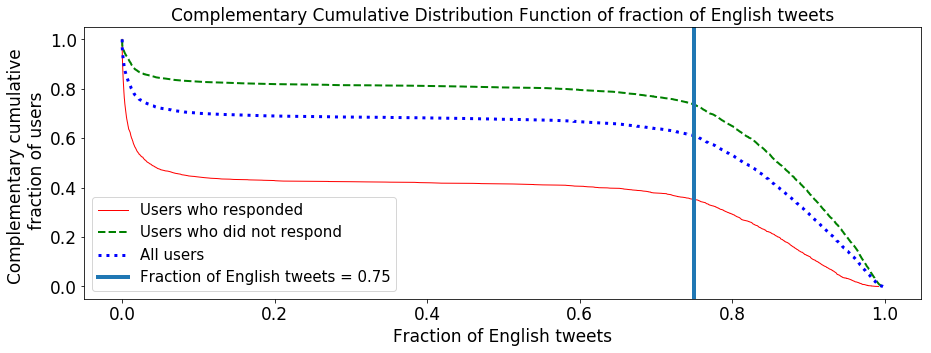}}
\caption{Complementary Cumulative Distribution Function of accounts whose fraction of English language tweets is greater than or equal to the x-axis. The x-axis shows the fraction of tweets in English, the y-axis shows the fraction of users in the three categories denoted in the legend whose fraction of English tweets is greater than the corresponding x-value.} 
\label{FractionEnglish_CCDF}
\end{figure}

\textbf{Language Analysis:} Next, a language analysis was performed using a language-detection Python library called langdetect \cite{langdetect}.  The two most predominant languages in the users' tweets were English and Russian. \figref{FractionEnglish_CCDF} shows the Complementary Cumulative Distribution Function (CCDF) of the fraction of tweets the users' published in English, as identified by langdetect \cite{langdetect}. As it can be inferred from the plot, any given user is most likely to either have a very low fraction (< 0.05) or an overwhelming majority (> 0.75) of English tweets, i.e the distribution of fraction of users' English tweets is a bimodal distribution. This is derived from the observation of the steep decrease in the CCDF at two points, at fractions close to 0.05 and 0.75, whereas for the x-values between these two steep decreases the CCDF is almost flat-like. Since this study aims at focusing on American Twitter users, we filter out users who do not have a significant fraction of their tweets in English. So based on the observed CCDF, only those users who had more than 75\% of their total tweets in English were selected to remain in the study. 2792 low follower count users and 2474 high follower count users were ultimately removed for having less than 75\% of their tweet history published in English.

\textbf{Removal For Insufficient Data:} Additionally, users whose first date of contact by an IRA account was prior to their first recorded tweet on file were removed (i.e., the IRA contact occurred before the account was made public). 219 low follower count user and 358 high follower count users were found have insufficient data and were removed, leaving a total of 1,839 users remaining in the low follower count user set, 578 of whom engaged back with an IRA account while the remaining 1,261 did not. These users formed the final responsive and unresponsive low follower count user sets, on which before and after analysis was performed. Of the high follower count users with 5,000 or greater follower counts, 3,056 users remained, 430 of whom engaged back with an IRA account and 2,626 did not. These two groups formed the responsive and unresponsive high follower count user groups respectively. 
Table~\ref{tab1} shows the number of tweets analysed in each category of users to conduct the before and the after analysis.

\textbf{Random Baseline of Twitter Users:} In order to ensure that user behavior observed in this work was sufficiently deviant from the greater population of Twitter during the years prior to the 2016 U.S. presidential election, a random set of Twitter users active during the same span of time around which this research focuses (2014-2016 primarily) has been included as a baseline.

To create this baseline set of random Twitter users, random number strings were generated to create random Twitter user ids. These were generated until four American users with accounts created in 2014 or earlier were found. These four users became the seeds used to create the remaining sample population. To form the remaining set of baseline users, we scraped the followers of these four seed accounts, and then continued to collect the \textit{k} followers of these followers of the original seed accounts, this processes continued endlessly until it was manually interrupted. In this manner, we utilized "snowball sampling" to grow our sample through continuously searching the follower lists of users encountered as a follower of a previously explored user \cite{10.2307/2237615}. 

This technique was used to find a total of over three million Twitter usernames. The list of Twitter users was randomly shuffled and the first 10,000 users underwent the same filtering process as the natural users who were contacted by the IRA (i.e., they had publicly accessible accounts, underwent bot analysis with Botometer and were shown to have 75\% or greater of their tweets published in English.) Ultimately, 8,023 users successfully exited the filtering process, had their complete history of available tweets collected, and were included as random baseline Twitter users in this research.

\begin{table}[b]
\caption{Size of the data-set used for the before-and-after analysis}
\renewcommand{\arraystretch}{1.2}
\footnotesize
\begin{tabular}{|c|c|c|c|c|}
\hline
\textbf{Category of users} & \textbf{Total count of users} & \textbf{Behavior} & \textbf{No. of users} & \textbf{No. of tweets}\\ \hline
\multirow{2}{*}{Low Follower Count}& \multirow{2}{*}{1,839} & Engaged back & 578 (31.4\%) & 19,548,890 \\ \cline{3-5} 
& & Did not engage back & 1,261 (68.6\%)& 33,960,883 \\ \hline

\multirow{2}{*}{High Follower Count}& \multirow{2}{*}{3,056} & Engaged back & 430 (14.07\%) & 29,246,645 \\ \cline{3-5} 
& & Did not engage back & 2,626 (85.95\%)& 148,780,230 \\ \hline

\end{tabular}

\label{tab1}
\end{table}


%% file: method.tex

\section{Method}


A before and after analysis was performed on the users contacted by IRA accounts. As each user was contacted on a potentially different date, time `zero' for each user in the study was assigned as the date of first IRA contact of that user, thus allowing us to aggregate before and after behaviors across different users contacted at different times.  To analyze the before IRA contact behavior, we considered for each user the 12-month time period preceding their time zero event, and similarly, to analyze the after IRA contact behavior, we considered for each user the 12-month time period following their time zero event. For each monthly score, a period of 31 days was used in place of a calendar month.


The before and after analysis monitored user behavior through six specific metrics: monthly user tweet count, monthly user average sentiment, number of monthly @realDonaldTrump mentions, number of monthly @HillaryClinton mentions, monthly average sentiment of tweets mentioning @realDonaldTrump and monthly average sentiment of tweets mentioning @HillaryClinton.

{\em Monthly Tweet Count} was determined for each user by assessing the total number of tweets published by a user within the bounds of a one-month period. This count includes tweets published by the Twitter user and replies to other Twitter accounts during that time period. We excluded favorite tweets, retweets of tweets authored by another Twitter account, or other Twitter activity from the monthly tweet count because of the limitations mentioned in the Data Collection Section. 

VADER Sentiment Analysis, a Python sentiment analysis tool built specifically to examine text written on social media, was used to return a sentiment score on the range [-1, 1] for each tweet authored by a given user within an allotted time span~\cite{vader}. A sentiment score of -1 indicates a highly negative sentiment while a score of 1 indicates a highly positive sentiment. These values were then averaged to form the {\em Monthly Sentiment} score for each user.

A Trump or Hillary mention is defined within this study as a tweet which directly tags either @realDonaldTrump or @HillaryClinton, respectively. These counts include tweets from users written in response to either @realDonaldTrump or @HillaryClinton. The sum of @realDonaldTrump and @HillaryClinton mentions over the assigned time period form the {\em Trump Mention Count} and {\em Hillary Mention Count}.

In order to examine the intersection of Trump or Hillary mentions and user sentiment, we used the VADER Sentiment Analysis on the tweets directed at @realDonaldTrump and @HillaryClinton. The results form the {\em Trump Mention Sentiment} and {\em Hillary Mention Sentiment} scores.

To assess the statistical significance of results from the before and after analysis of user tweeting behaviors, the Wilcoxon hypothesis test was used \cite{wilcoxon}, which is a non-parametric alternative to the dependent samples t-test. We use this alternative because the distribution of our sample data does not approximate a normal distribution. Also, since the before and after samples are not independent of each other, the Mann-Whitney U-test cannot be used. The assumptions of the Wilcoxon tests are
(a) the samples compared are paired, (b) each pair is chosen randomly and independently, and (c) the data are measured on at least an interval scale when within-pair differences are calculated to perform the test.
The samples we use to perform the Wilcoxon hypothesis tests satisfy all these three assumptions. Since not all users may have 12 months of activity on Twitter before they were first contacted by the IRA, for the hypothesis tests we consider only a subset of users who were active on Twitter for at least 6 months prior to their interaction with the IRAs. This also removes the users who were very new to the platform. All the hypothesis tests were done at a 0.05 level of significance.

Note that for the random baseline users, there are no \textit{before-and-after} values because these users did not have contacts with the IRAs. Hence, we do not use the Wilcoxon hypothesis test for comparing the users the IRA contacted with the 8,023 random Twitter users. Instead, we use the Kolmogorov–Smirnov hypothesis test (KS test) \cite{massey1951kolmogorov}. The KS test can quantify whether two samples come from the same distribution or not. Also, the KS test is applicable to our data because it is a non-parametric test and hence does not assume that the samples to be derived from Gaussian distributions. We run the KS test at 0.05 level of significance to verify whether the behaviours of the random baseline users are significantly different from the users the IRAs interacted with.



%% file: results.tex

\section{Results}
\label{sec_res}
\subsection{User Behavior}
A before and after analysis was performed on the behavior of Twitter users contacted by IRA accounts. 
Using the prior definitions of low follower count users and high follower count users, as well as responsive and non-responsive users, four groups were analyzed, namely responsive low follower count users, unresponsive low follower count users, responsive high follower count users, unresponsive high follower count users. Plots have behavior for the year prior to and following IRA contact displayed, whereas significance testing, means, medians and other descriptive data are determined using only six months before and after IRA contact. Due to the large variance of behavior among user groups, we have chosen to illustrate the means of this data with bar graphs in Figures 4-15, 19.

\begin{figure}[t]
\centerline{\includegraphics[width=350pt]{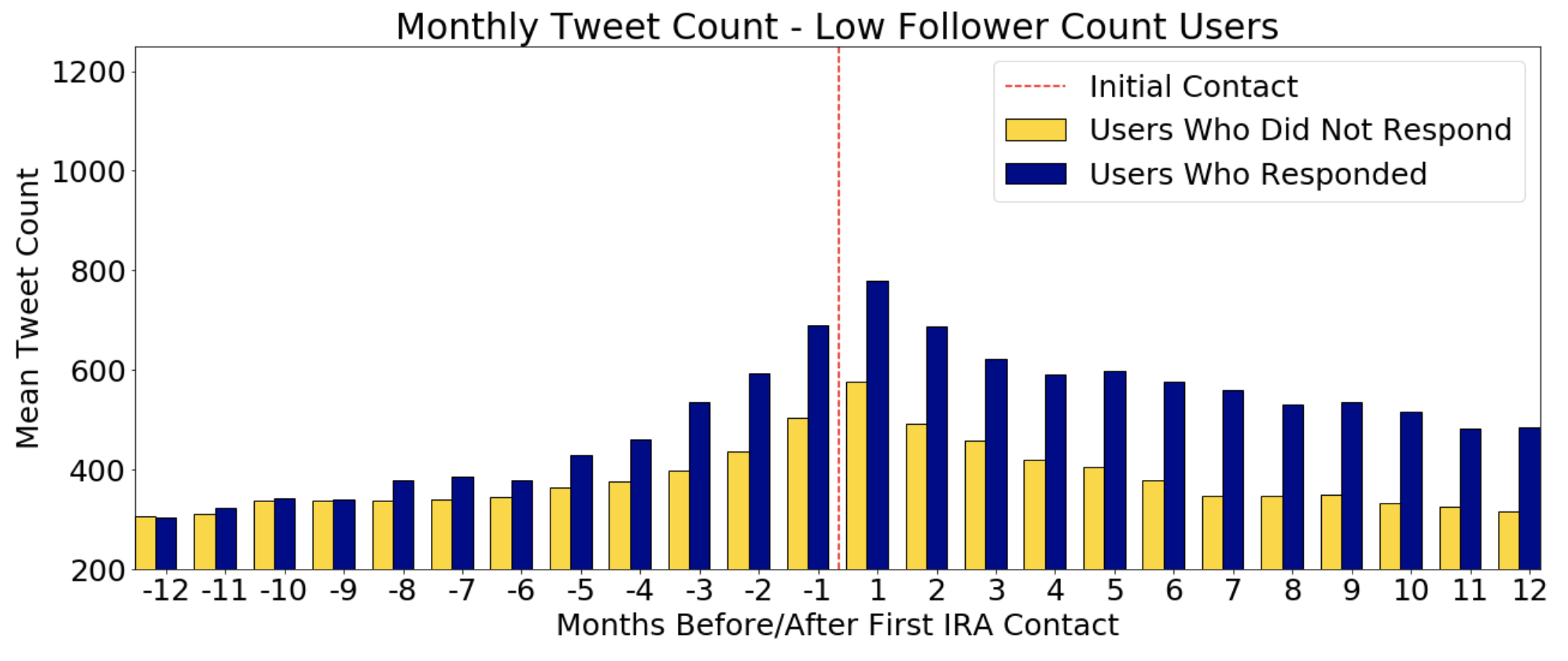}}
\caption{Total monthly tweet volumes have changed in a statistically significant way after first contact in both responsive and unresponsive low follower count user groups. Both subsets had local maximum tweet counts in the month following IRA contact.} 
\label{noninf_tweetcount}
\end{figure}

\begin{figure}[t]
\centerline{\includegraphics[width=350pt]{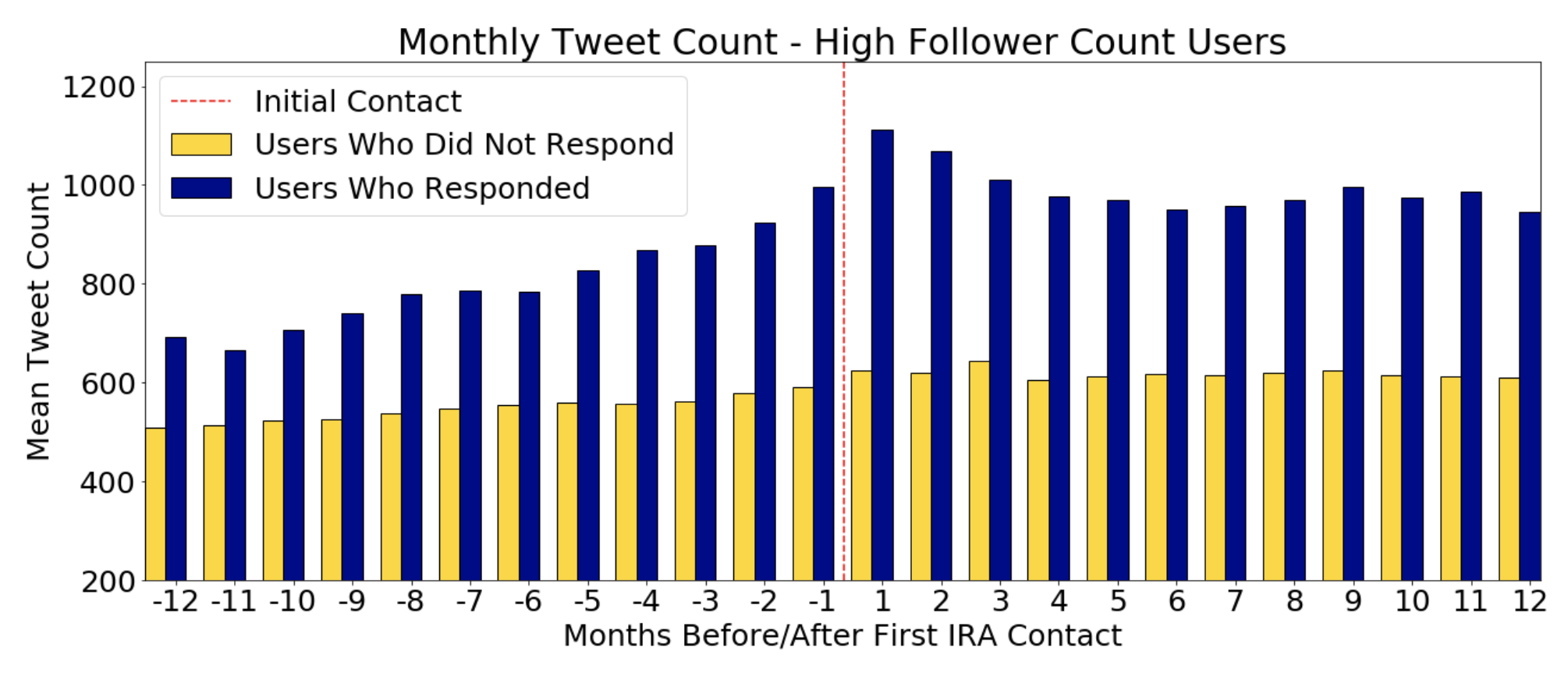}}
\caption{Total monthly tweet volumes have increased in a statistically significant manner after first contact in both responsive and unresponsive high follower count user groups. Only responsive high follower count users are shown to have had a local maximum tweet count occur in the month directly following initial IRA engagement.} 
\label{inf_tweetcount}
\end{figure}

\textbf{Monthly Tweet Frequency:} 
\textit{Users across all groups were found to have significantly changed tweeting frequency in the months following initial IRA contact} (See \figref{noninf_tweetcount} and \figref{inf_tweetcount}). Responsive low follower count users had an average increase in tweet volume of 26.02\% (430.02 average tweets/month before, 579.89 tweets/month after) following first IRA contact (\textit{P} = $3.56 \times 10^{-14}$) and unresponsive low follower count users had an average increase in tweet volume of 15.43\% (365.99 tweets/month before, 395.38 tweets/month after) (\textit{P} = $1.23 \times 10^{-5}$). High follower count users saw an average increase of 15.96\% (804.24 tweets/month before, 993.09 tweets/month after) for responsive users (\textit{P} = $1.02 \times 10^{-5}$) and 9.83\% (546.48 tweets/month before, 618.39 tweets/month after) for unresponsive users (\textit{P} = $3.83 \times 10^{-12}$). Within both the responsive and unresponsive low follower count users as well as the responsive high follower count users, a local maximum is seen as having occurred in the month following first IRA contact, however this trend is not followed in the the unresponsive high follower count user subset. It is shown here that the responsive users within both the low follower count and high follower count sets have clearly sustained higher mean tweets in the year following IRA interaction. While the unresponsive low follower count users also show a peak during the month following IRA interaction, their tweet counts descend to reflect counts similar to those found before IRA contact.

\begin{figure}[t]
\centerline{\includegraphics[width=350pt]{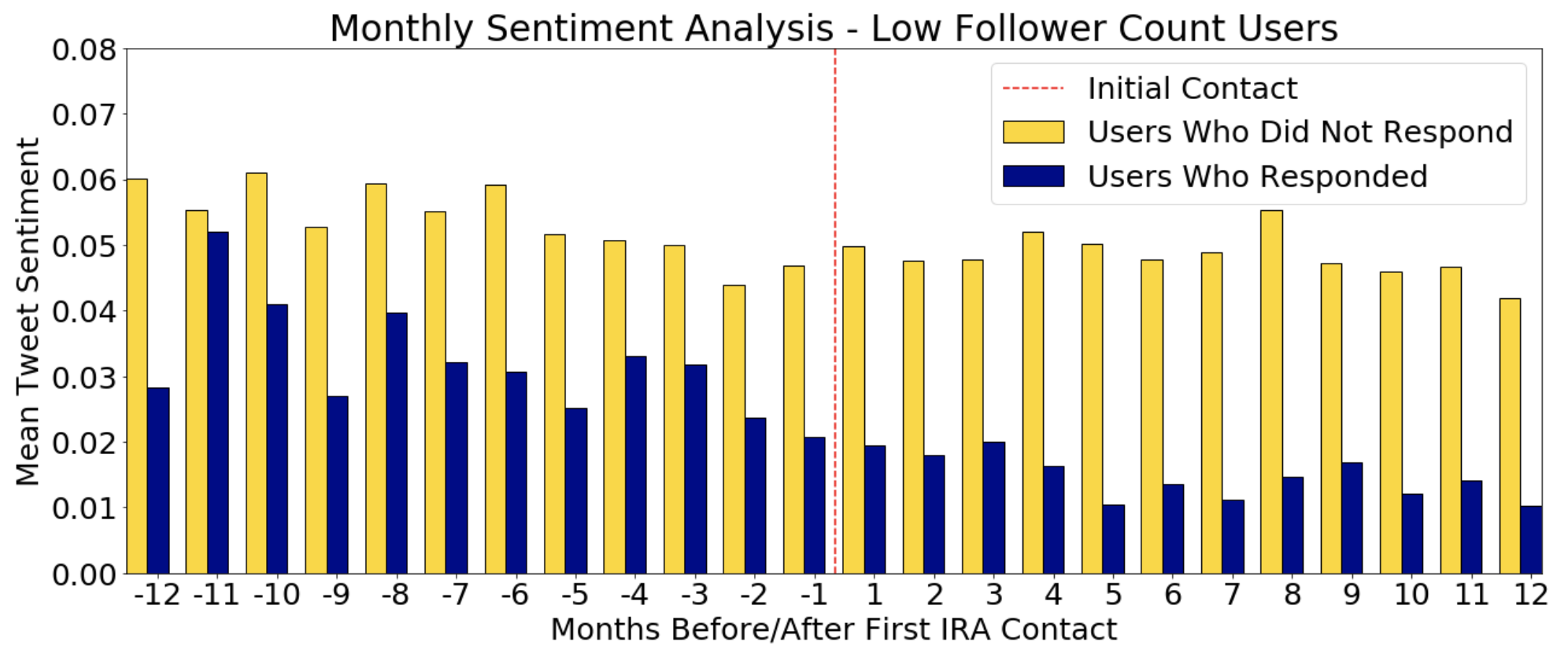}}
\caption{Average monthly sentiment significantly changed only for low follower count users which responded back to an IRA accounts. Users who chose not to engage back with IRA accounts did not show significant change at the 0.05 level of significance in sentiment in the months following contact.} 
\label{non_sent}
\end{figure}

\begin{figure}[t]
\centerline{\includegraphics[width=350pt]{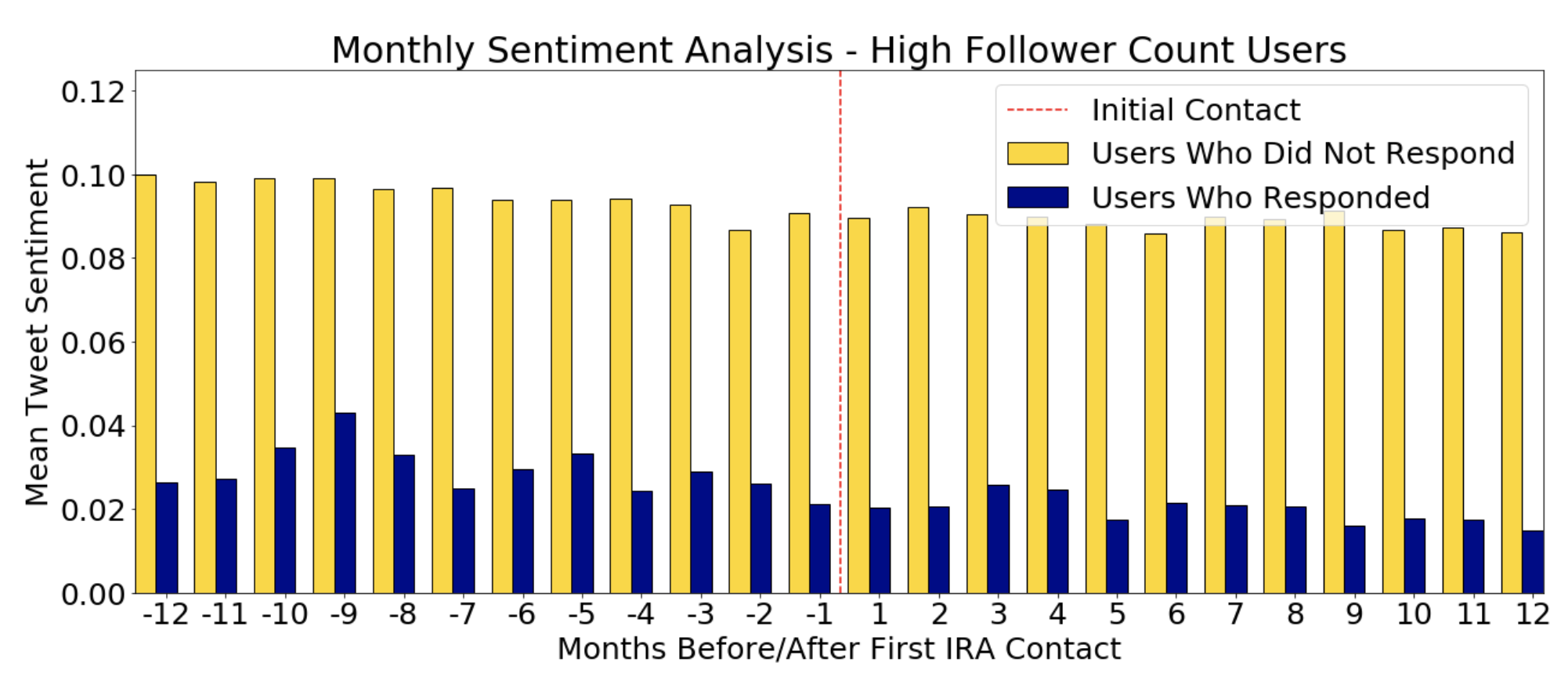}}
\caption{Average monthly sentiment significantly changed for both high follower count user groups, responsive and unresponsive.} 
\label{inf_sent}
\end{figure}

\textbf{Monthly Tweet Sentiment:} \textit{Overall mean tweet sentiment by month was shown to increase in negativity in a statistically significant manner for responsive low follower count users, and both responsive and unresponsive high follower count users} (See \figref{non_sent} and \figref{inf_sent}). For responsive low follower count users, mean sentiment prior to IRA contact rested at 0.023 and fell to 0.015 after and this increased negativity was shown to be significant (\textit{P} = $0.0001$), whereas unresponsive low follower count users had a mean sentiment of 0.049 before IRA contact and 0.05 after, and this change was not significant (\textit{P} = $0.62$). For high follower count users, mean sentiment for responsive users was 0.022 before contact and 0.019 after (\textit{P} = $0.024$), while for the unresponsive ones the mean sentiment was 0.092 before IRA contact and 0.088 after (\textit{P} = $0.01$). Each of these changes was found to be significant. Responsive low follower count users became notably more negative in sentiment after the IRA contact, while responsive high follower count users tended to be more negative in general than their unresponsive counterparts.

\begin{figure}[t]
\centerline{\includegraphics[width=350pt]{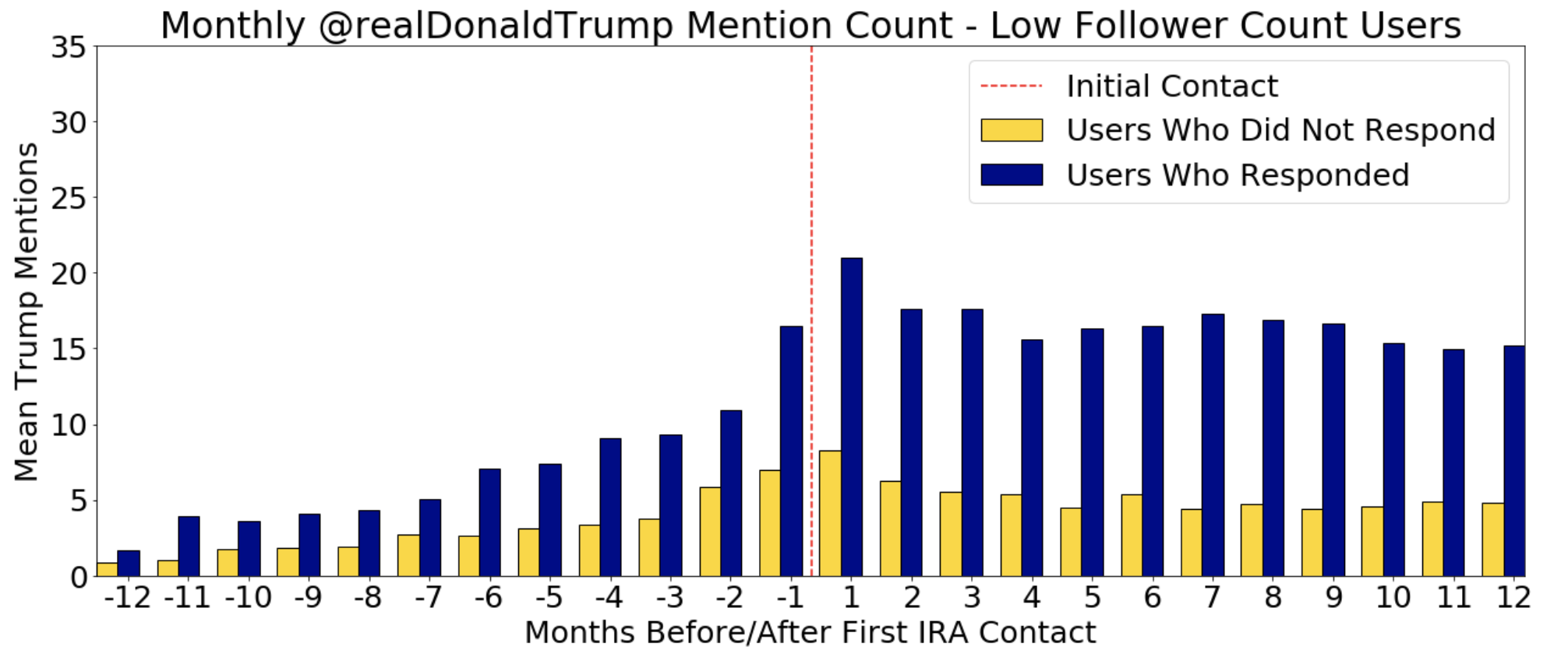}}
\caption{The monthly @realDonaldTrump mention counts for the low follower count users groups, those which responded to an IRA account and those which did not.} 
\label{non_trump}
\end{figure}

\begin{figure}[t]
\centerline{\includegraphics[width=350pt]{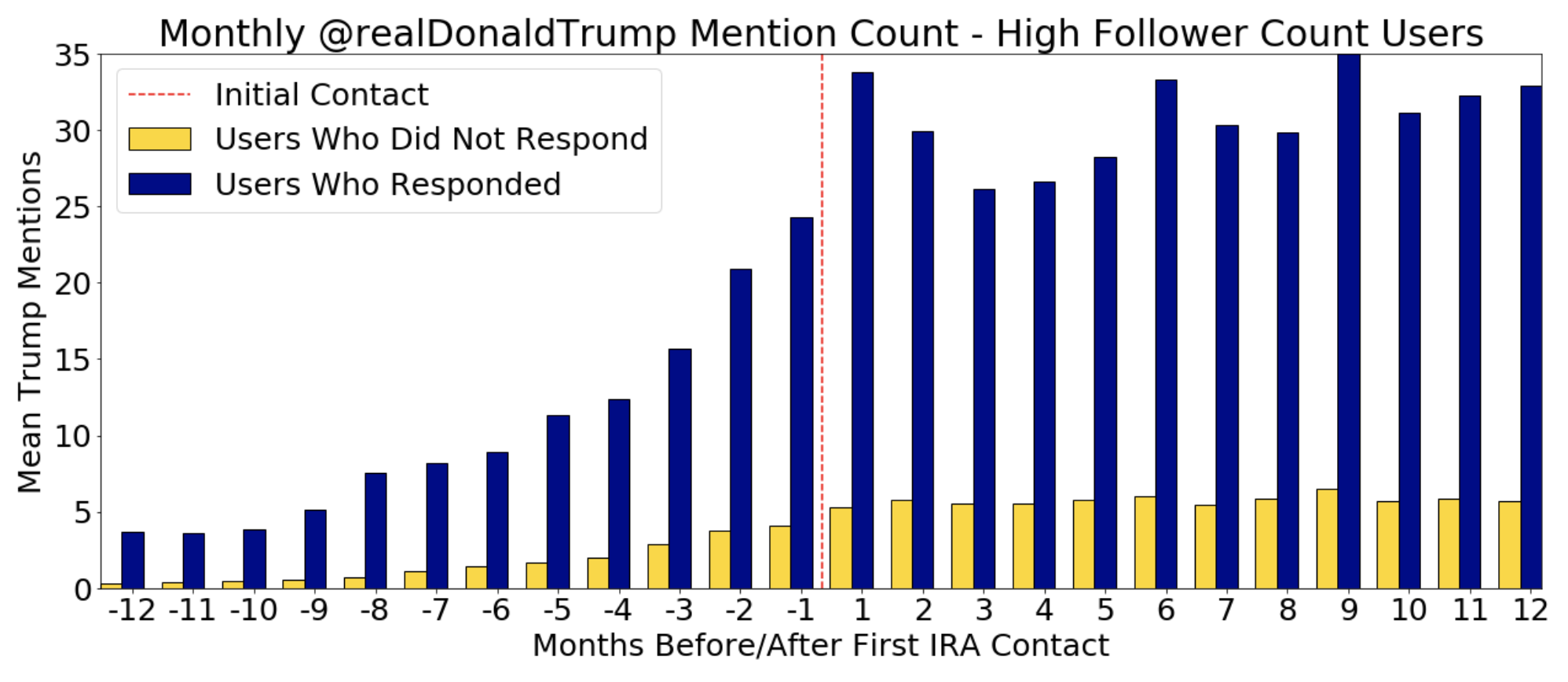}}
\caption{The monthly @realDonaldTrump mention counts for the high follower count user groups, those which responded to an IRA account and those which did not. The responsive high follower count accounts mentioned @realDonaldTrump more regularly than any other group.} 
\label{inf_trump}
\end{figure}

\textbf{@realDonaldTrump Mentions:} \textit{The number of monthly mentions of the @realDonaldTrump account significantly increased for all user segments} (See \figref{non_trump} and \figref{inf_trump}). For low follower count users, mentioning of the @realDonaldTrump handle increased by 65.51\% (from 6.92 average mentions/month before to 16.75 mentions/month after) for responsive users (\textit{P} = $9.12 x 10^{-18}$) and by 27.93\% (from 2.96 mentions/month before to 5.27 mentions/month after) for unresponsive users (\textit{P} = $1.89 x 10^{-14}$). Among high follower count users, @realDonaldTrump was mentioned 94.72\% more (from 10.46 mentions/month before to 30.79 mentions/month after) after IRA contact by responsive users (\textit{P} = $9.14 x 10^{-23}$) and 145.82\% more (from 1.62 mentions/month before to 5.74 mentions/month after) by unresponsive users (\textit{P} = $1.77 x 10^{-88}$). In the responsive users groups, low and high follower count users, the number of @realDonaldTrump mentions dwarfs the unresponsive user groups in their respective categories. Responsive high follower count users were found to have tagged @realDonaldTrump with more regularity than any other user set.

\begin{figure}[t]
\centerline{\includegraphics[width=350pt]{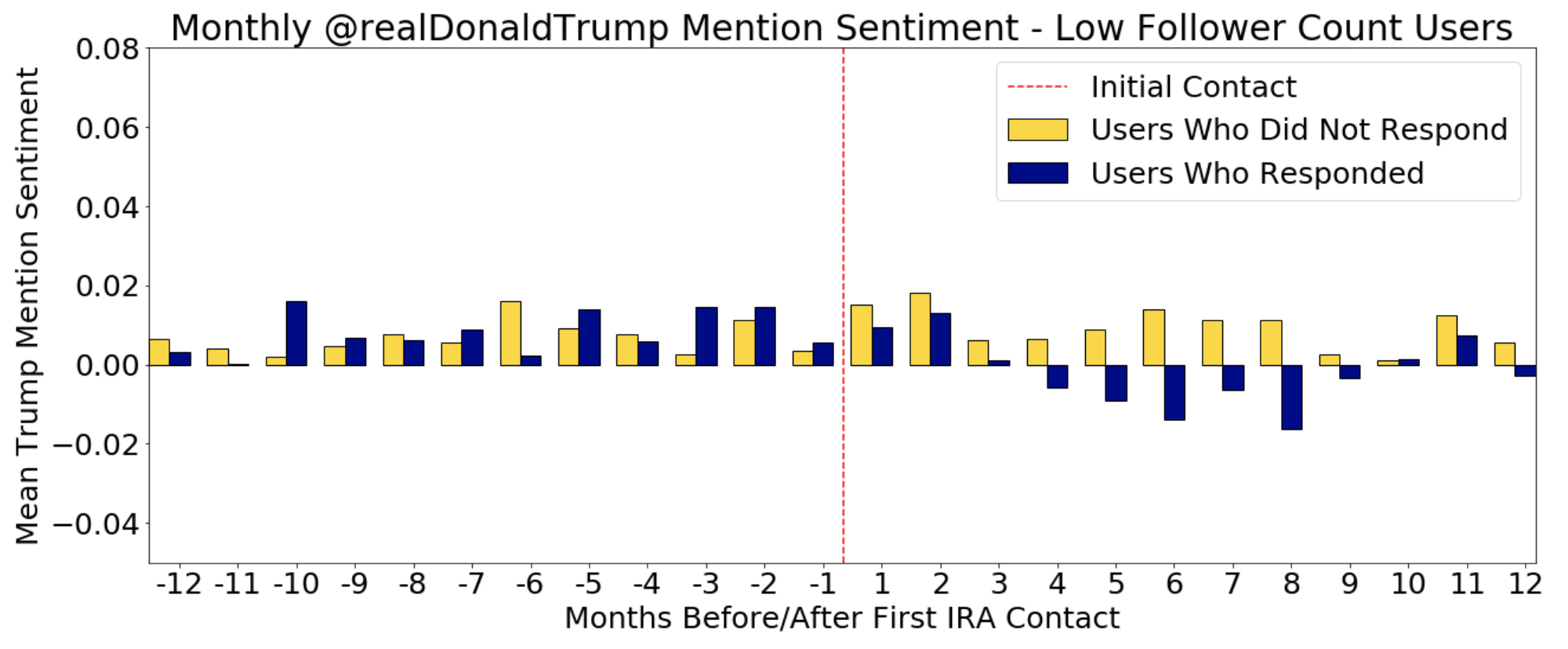}}
\caption{No significant change was detected in the sentiment of @realDonaldTrump tweets by contacted users for either the responsive or unresponsive low follower count users, despite the dip into an overall negative sentiment seen in responsive low follower count users.} 
\label{non_trump_sent}
\end{figure}

\begin{figure}[t]
\centerline{\includegraphics[width=350pt]{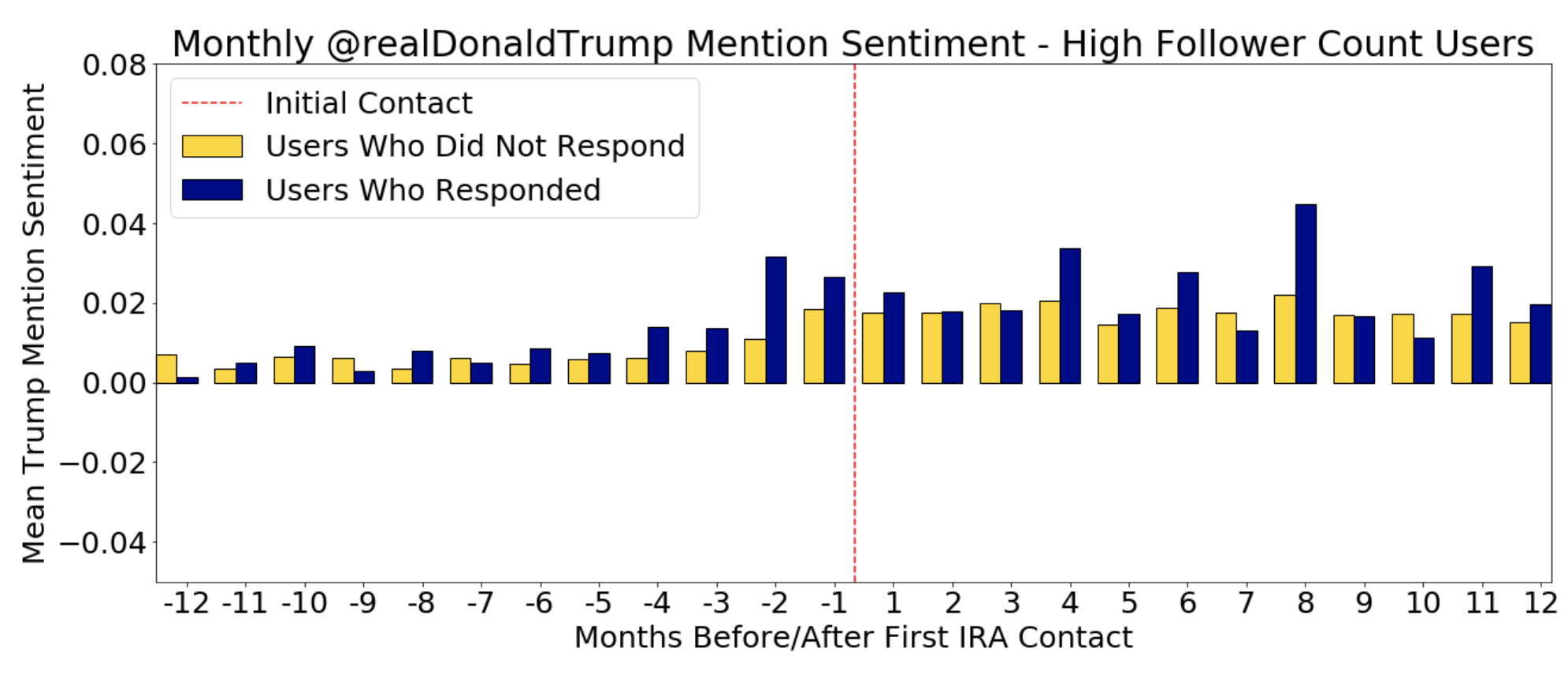}}
\caption{Sentiment of @realDonalTrump tweets by contacted high follower count users was shown to have significantly changed only for unresponsive high follower count users.  For responsive high follower count users, no significant change was detected.} 
\label{inf_trump_sent}
\end{figure}

Looking at the intersection of user sentiment and @realDonaldTrump mentions, a significant change was observed only within the unresponsive high follower count user set (See \figref{non_trump_sent} and \figref{inf_trump_sent}). While responsive low follower count users are seen to become increasingly negative in the subsequent months following IRA contact (going from a mean sentiment of 0.002 before and -0.015 after IRA contact), this was not found to be a statistically significant change (\textit{P} = $0.068$). A random analysis of this tweet set indicated that while tweet sentiment did grow more negative, the negativity was most often targeting the opponents of @realDonaldTrump, often stemming from tweets replying to @realDonaldTrump himself. Some examples of this phenomenon are: "@readDonaldTrump We know Obama had Trump Tower wire tapped no doubt at all keep tweeting", "@foxnews [...] @realDonaldTrump It's not bigotry to keep terrorists out of this country." and "@nytimes @realDonaldTrump [...] Everyone knows media \& NYT is dishonest. That is why you are popular, you tell them to their face." (some identifying usernames have been redacted from these examples). It appeared that the sentiment change in @realDonaldTrump tweets may have indicated a growing desire to engage Trump's perceived adversaries than an increasingly negative sentiment for Trump himself. Unresponsive low follower count users sentiment decreased from 0.015 before IRA contact to 0.009 after, this change was also not shown to be significant (\textit{P} = $0.318$). The decline into negative sentiment of @realDonaldTrump mentions seen in the unresponsive low follower count user group is not seen within the high follower count subset.  Unresponsive high follower count users appeared to conversely grow more positive in their tweets regarding @realDonaldTrump, with a before mean sentiment of 0.019 and an after sentiment of 0.029, this change was shown to have significantly increased (\textit{P} = $0.003$). Responsive high follower count users had a mean sentiment of @realDonaldTrump mentions of 0.031 before IRA contact and 0.013 after, this change was not proven statistically significant (\textit{P} = $0.32$).

\begin{figure}[t]
\centerline{\includegraphics[width=350pt]{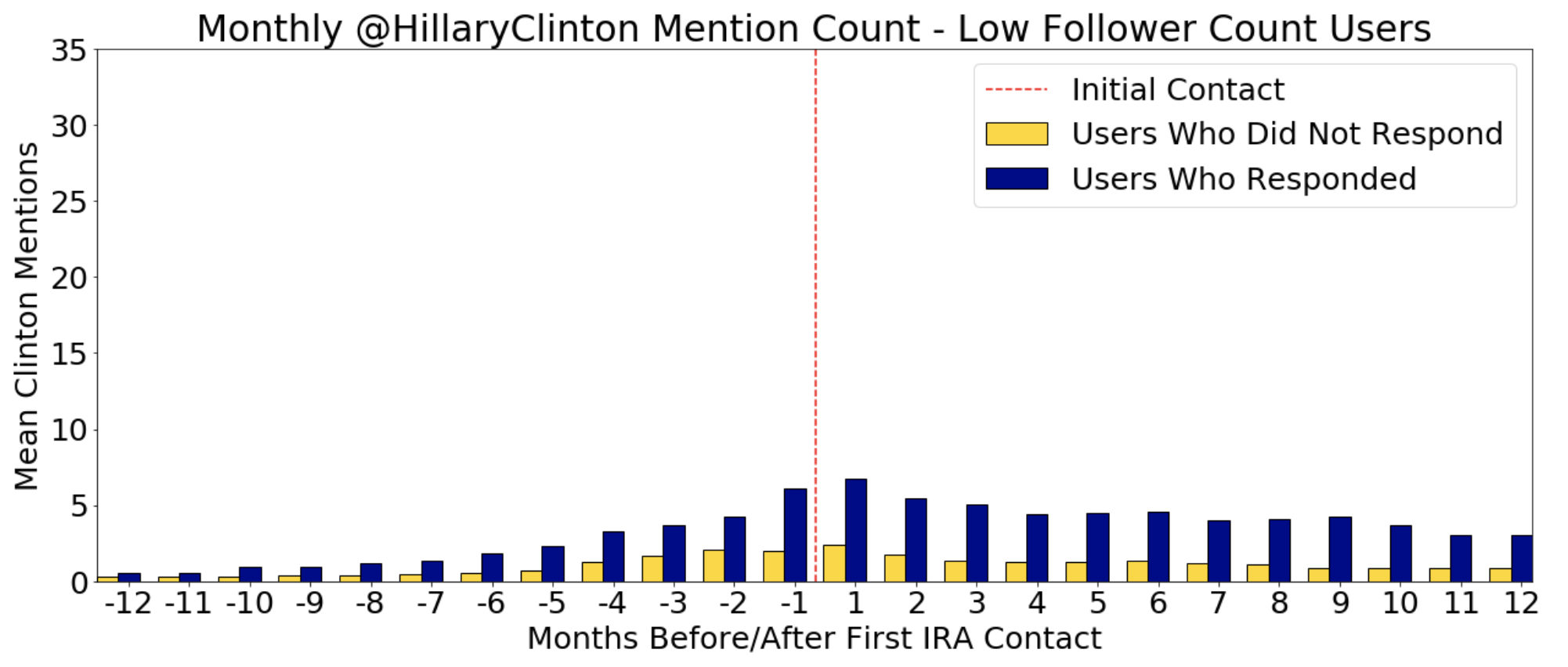}}
\caption{The change in monthly @HillaryClinton mention counts are significant among each of the low follower count user groups, responsive and unresponsive.} 
\label{non_hillary}
\end{figure}

\begin{figure}[t]
\centerline{\includegraphics[width=350pt]{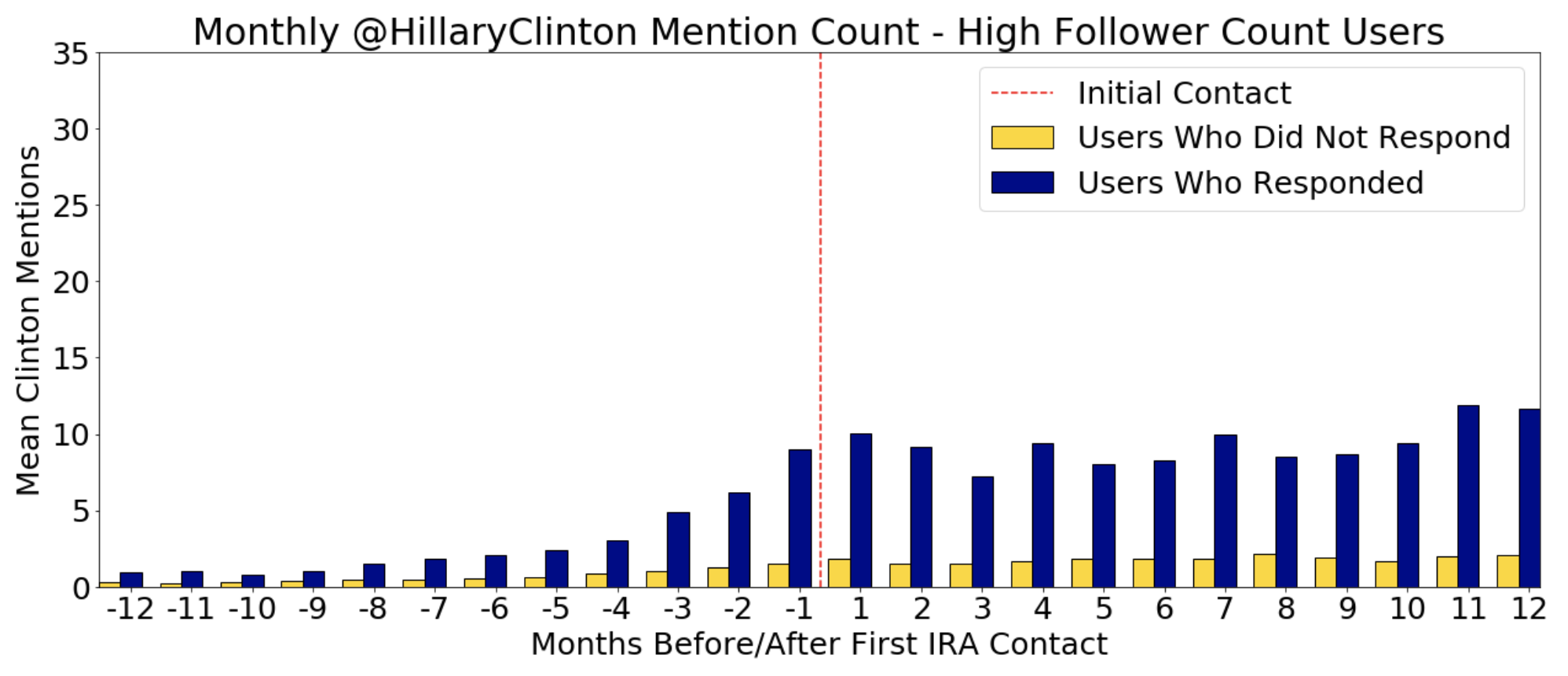}}
\caption{The change in monthly @HillaryClinton mention counts are significant among each of the high follower count user subsets, those which responded to an IRA account and those which did not.} 
\label{inf_hillary}
\end{figure}

\textbf{@HillaryClinton Mentions:} \textit{There is a statistically significant increase in the number of monthly mentions of @HillaryClinton within all user groups} (See \figref{non_hillary} and \figref{inf_hillary}). For low follower count users, responsive users mentioned @HillaryClinton with 55.44\% more (from 2.24 mentions/month before to 4.41 mentions/month after) regularity after IRA contact (\textit{P} = $2.23 x 10^{-9}$), and unresponsive users mentioned the handle 14.75\% more (from .86 mentions/month before to 1.28 mentions/month after) often (\textit{P} = $2.7 x 10^{-13}$), each of these increases was found to be significant. For high follower count accounts, there was an increase of 101.62\%  (from 2.89 mentions/month before to 9.35 mentions/month after) and 67.4\% (from .68 mentions/month before to 1.82 mentions/month after) in @HillaryClinton mentions for responsive (\textit{P} = $2.7 x 10^{-13}$) and unresponsive users (\textit{P} = $1.82 x 10^{-34}$) respectively, and each of these increases was found to be statistically significant. While all groups showed increased mentions of @HillaryClinton, responsive users from each the high or low follower count user sets clearly sustained a larger number of mentions in the year after first IRA contact, following the pattern that @realDonaldTrump mention counts exhibited to a far lesser degree. Again responsive high follower count users are found to have mentioned @HillaryClinton more than any other user group, though their use of the @realDonaldTrump handle triples their use of @HillaryClinton for most months.

\begin{figure}[t]
\centerline{\includegraphics[width=350pt]{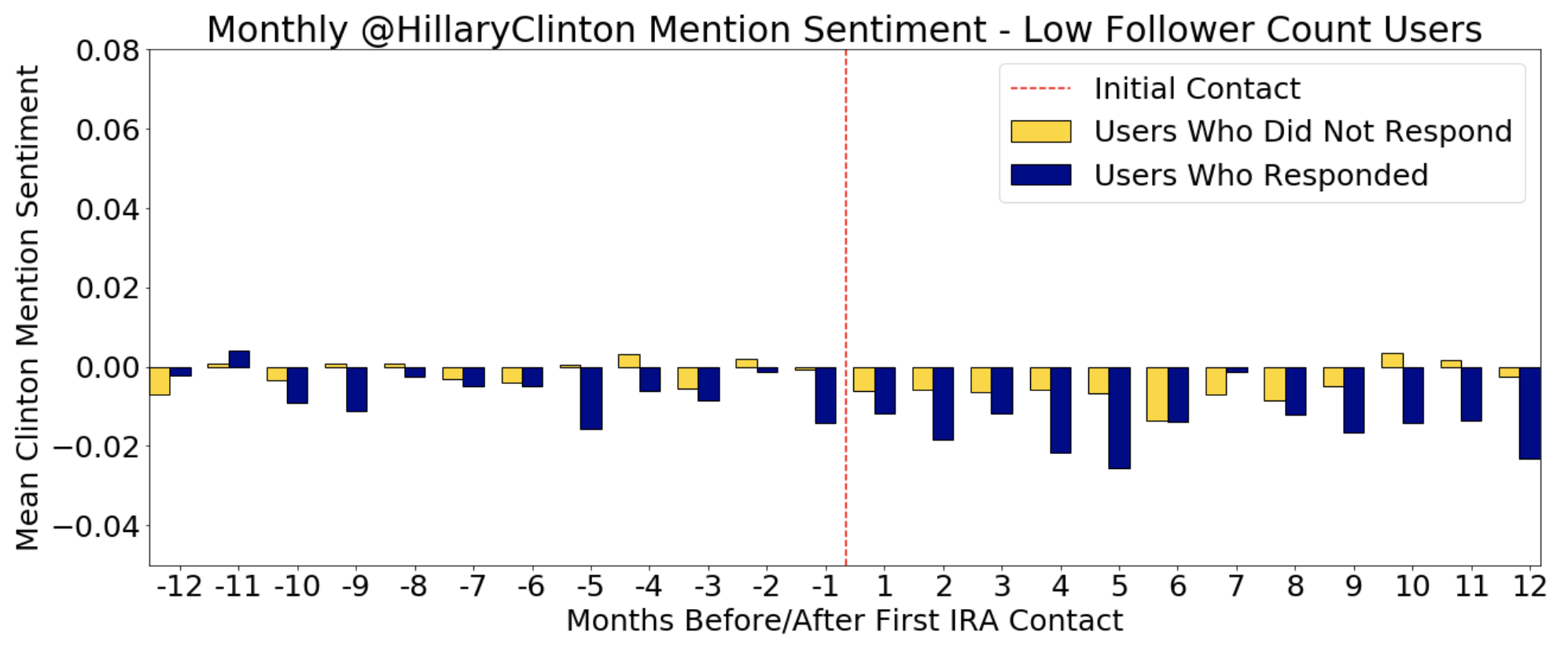}}
\caption{Sentiment has significantly changed only for unresponsive low follower count users mentioning @HillaryClinton. There was no significant change for responsive low follower count users.}
\label{non_hillary_sent}
\end{figure}

\begin{figure}[t]
\centerline{\includegraphics[width=350pt]{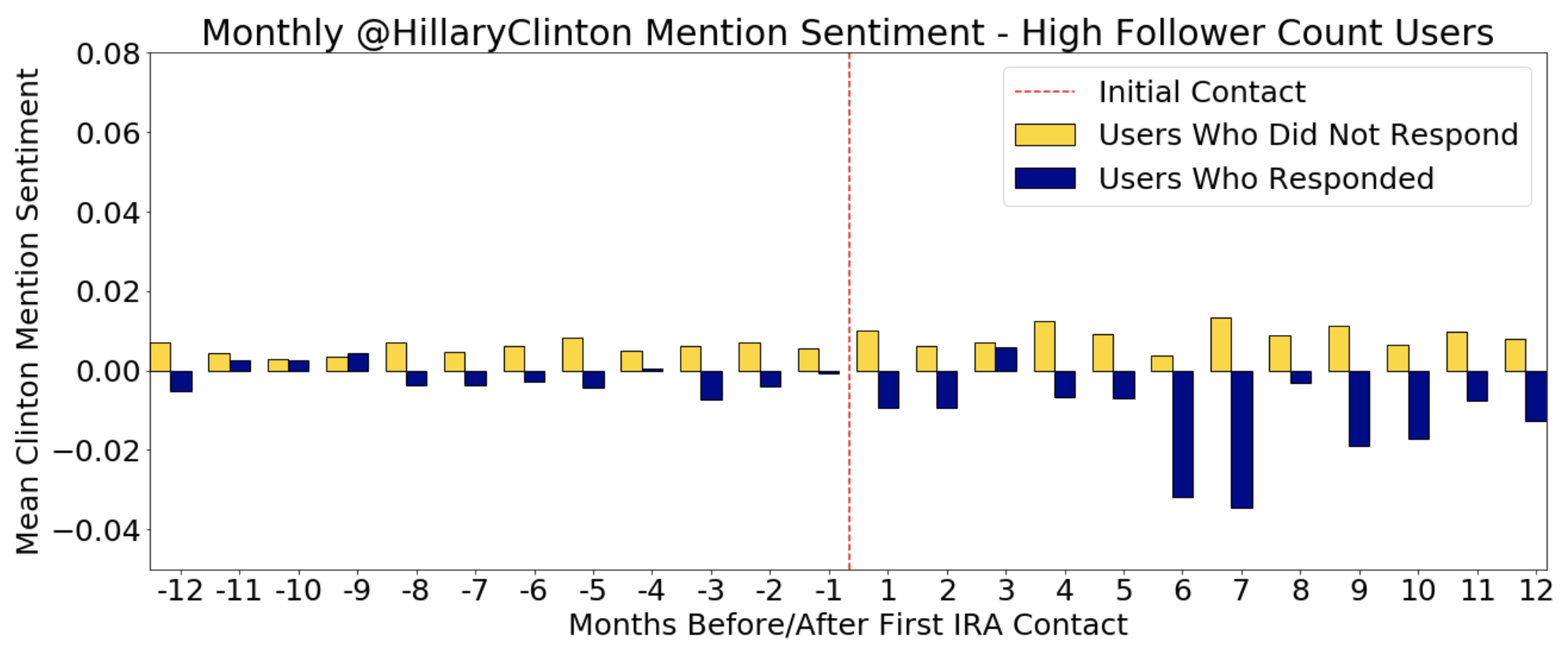}}
\caption{Sentiment has significantly changed for responsive high follower count users mentioning @HillaryClinton. There is no significant change for unresponsive high follower count users.}
\label{inf_hillary_sent}
\end{figure}

There is a statistically significant increase in negativity of user sentiment for the @HillaryClinton mentions only within the unresponsive low follower count users and the responsive high follower count users (See \figref{non_hillary_sent} and \figref{inf_hillary_sent}). Responsive low follower count users have a mean sentiment of @HillaryClinton mentions of -0.017 before IRA contact and -0.021 after, a change which was not found to be significant (\textit{P} = $0.322$). Unresponsive low follower count users have a before mean of 0.002 and an after mean of -0.001 and this change was shown to be statistically significant (\textit{P} = $0.025$). High follow count users' mean sentiment changes from -0.003 to -0.025 for responsive users (\textit{P} = $0.39$) and 0.018 to 0.017 for unresponsive users (\textit{P} = $0.003$), with only the change in responsive users shown to be significant. Possibly too few data points exist to show significant change within this set, as @HillaryClinton was not mentioned as often as @realDonaldTrump.

\subsection{IRA to User Contact Points}

\begin{figure}[t]
\centerline{\includegraphics[width=350pt]{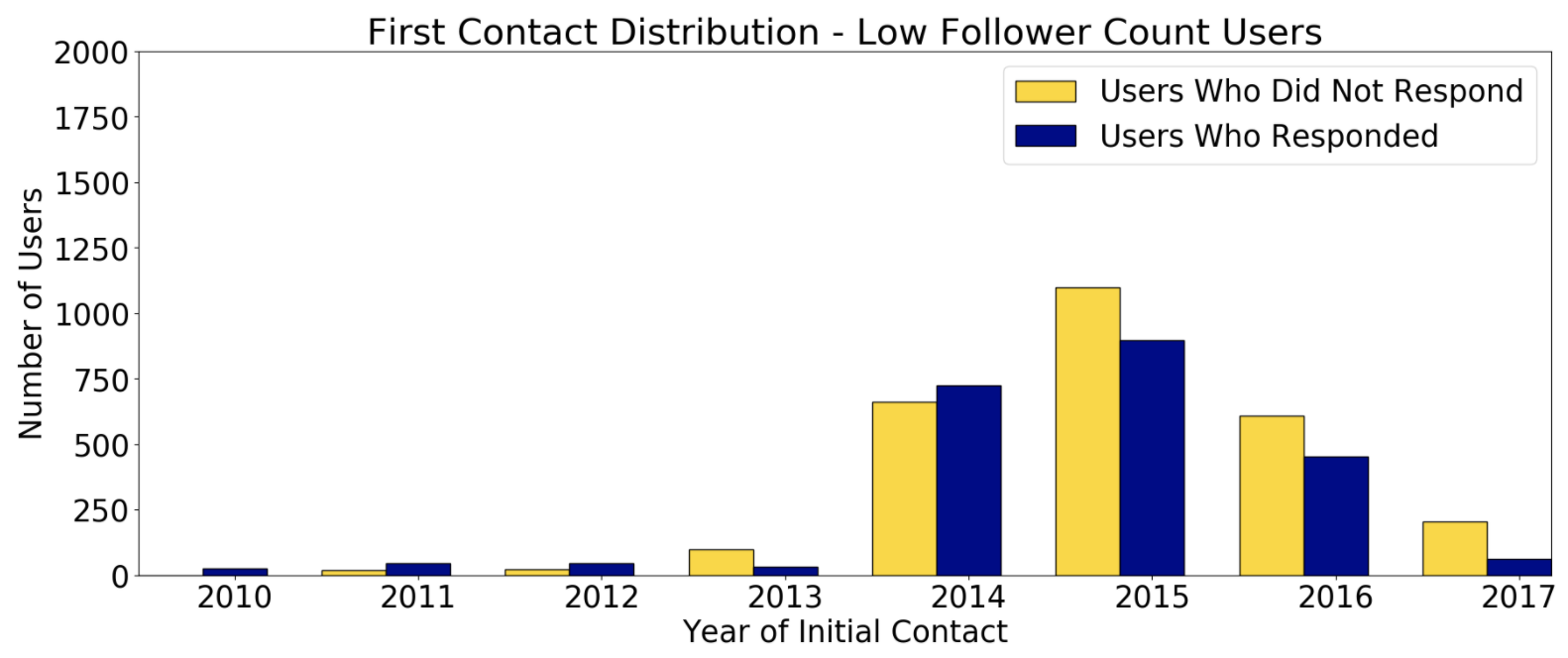}}
\caption{The number of IRA outreach to general Twitter users increases until the spike in users reaches a maximum in 2015, the year proceeding the 2016 U.S. Presidential election. After the diminished activity in 2017 no further users were contacted by the IRA accounts housed within the data set of this study.} 
\label{first_contact_noninf}
\end{figure}

\begin{figure}[t]
\centerline{\includegraphics[width=350pt]{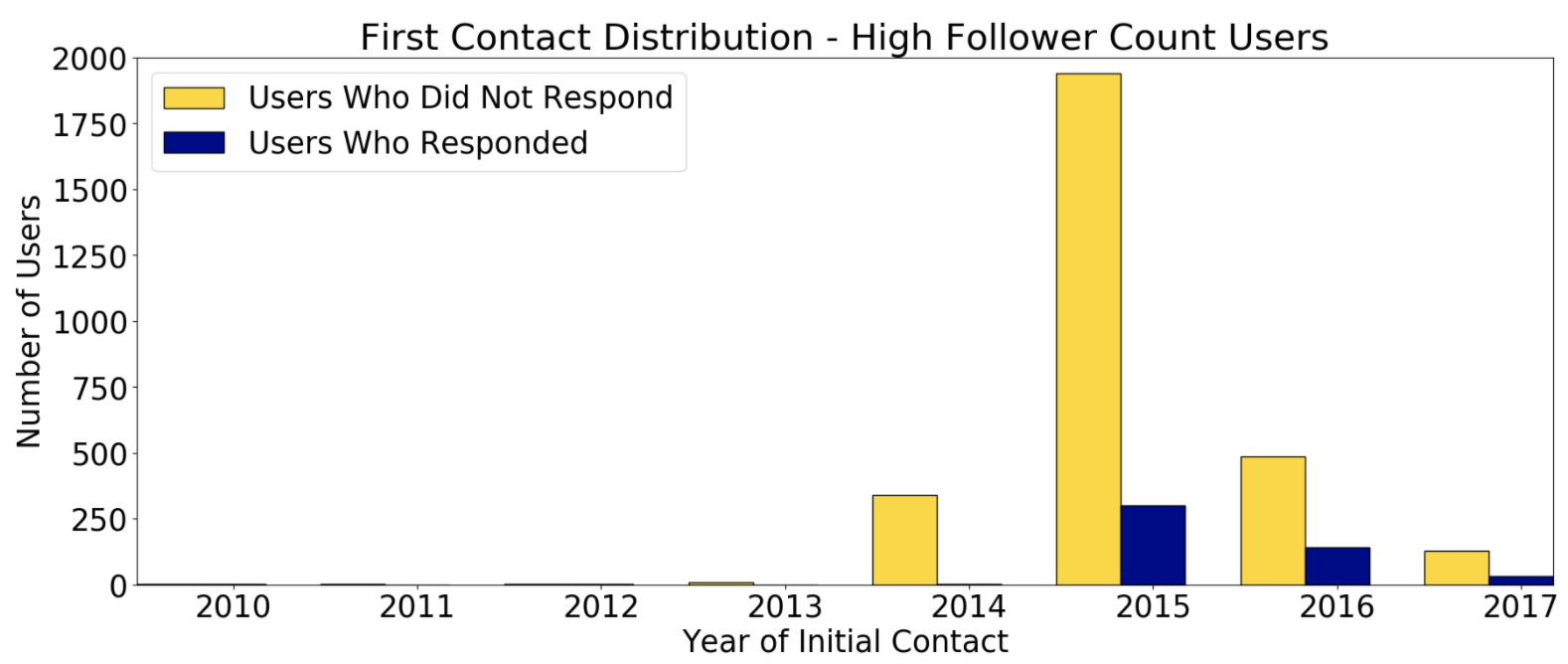}}

\caption{The majority of high follower count accounts were contacted by the IRA first in 2015, and no further initial contacts were found after 2017. Far fewer high follower count accounts were reached within years prior to 2014.} 

\label{first_contact_inf}
\end{figure}

The following data has been presented in order to contextualize the strategies and the timeline of events surrounding IRA's misinformation campaign (\figref{first_contact_noninf} and \figref{first_contact_inf})

The IRA accounts listed within the data were created as early as 2009, though the majority of accounts were registered with Twitter in 2014. Accounts continued to be created through 2016. Examination of the distribution of first contact dates, i.e., the date upon which a user was first mentioned by an IRA account, showed that while 2014 saw the first large wave of users being contacted by IRA accounts, the most extensive campaign to reach Twitter users initiated contact within 2015. Activity in 2016 decreased to mirror the number of users in 2014, before further dropping off to the final year of actively reaching new Twitter users in 2017 (See  \figref{first_contact_noninf} and \figref{first_contact_inf}).

\begin{figure}[t]
\centerline{\includegraphics[width=350pt]{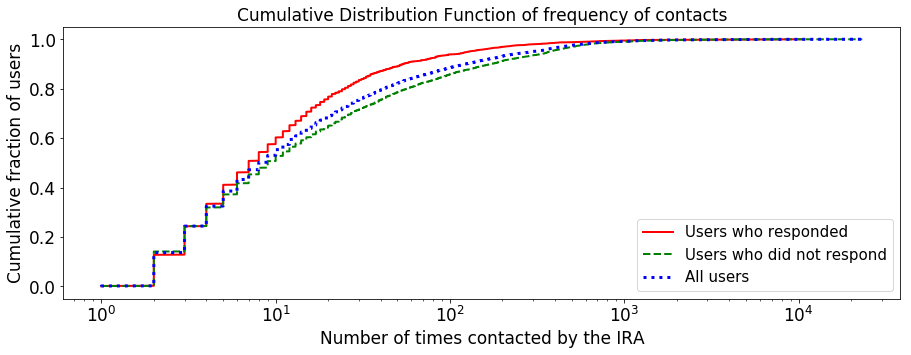}}

\caption{Cumulative Distribution Function of number of times the users got contacted by the IRA. Here the user set is collective of users with less than 5000 follower count as well as those greater than or equal to 5000. The x-axis shows the frequency of contact by the IRA in a log-scale and the y-axis shows the fraction of users who have been contacted by the IRA the number of times less than or equal to the corresponding x-value.} 

\label{CDF_contact}
\end{figure}

The mean number of contact points from an IRA account to a user is highly skewed due to the number of contact points with a relatively small number of accounts, such as @realDonaldTrump. The Cumulative Distribution Function of the frequency of IRA contacts is shown in \figref{CDF_contact} over a log-scale. For users who responded back, for those who did not, and for all the users collectively, the median number of times an user was contacted by the IRA was 7, 9 and 8 respectively.

The average time span from the first to the last contact with an IRA account across all users, is 503.67 days, well over an entire year. Within the low follower count user group we see that the mean span of contact is 355.23 days (449.08 mean for responsive low follower count users and 276.12 for unresponsive low follower count users). For high follower count users, the average span of contact is 636.32 days (629.23 day mean for responsive high follower count users and 642.85 day mean for unresponsive users).

\subsection{Random Twitter User Baseline}

\begin{figure}[b]
\centering
\begin{subfigure}{.5\textwidth}
    \centering
    \includegraphics[width=1.0\textwidth]{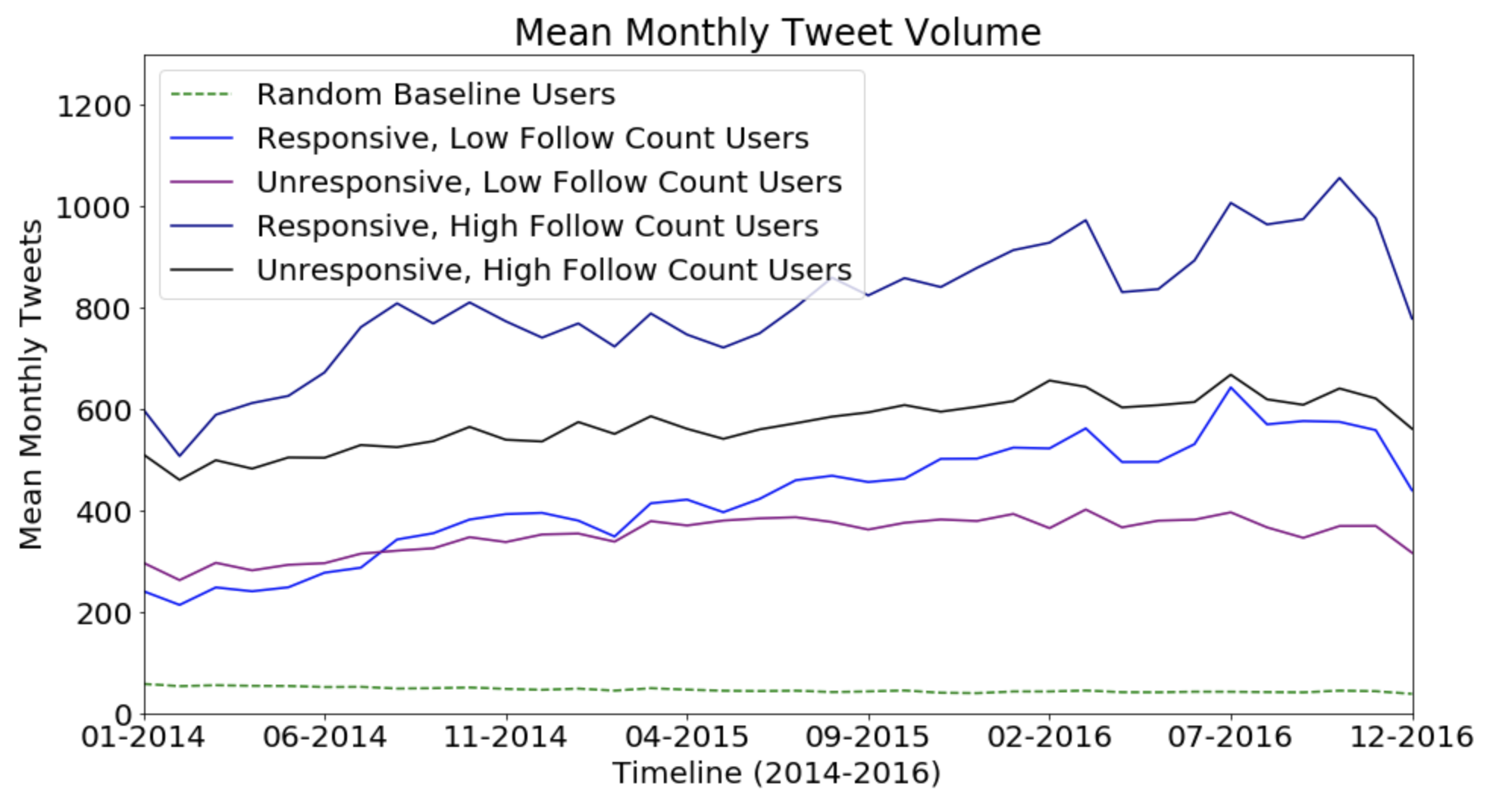}
    \caption{}
\end{subfigure}%
\begin{subfigure}{.5\textwidth}
    \centering
    \includegraphics[width=1.0\textwidth]{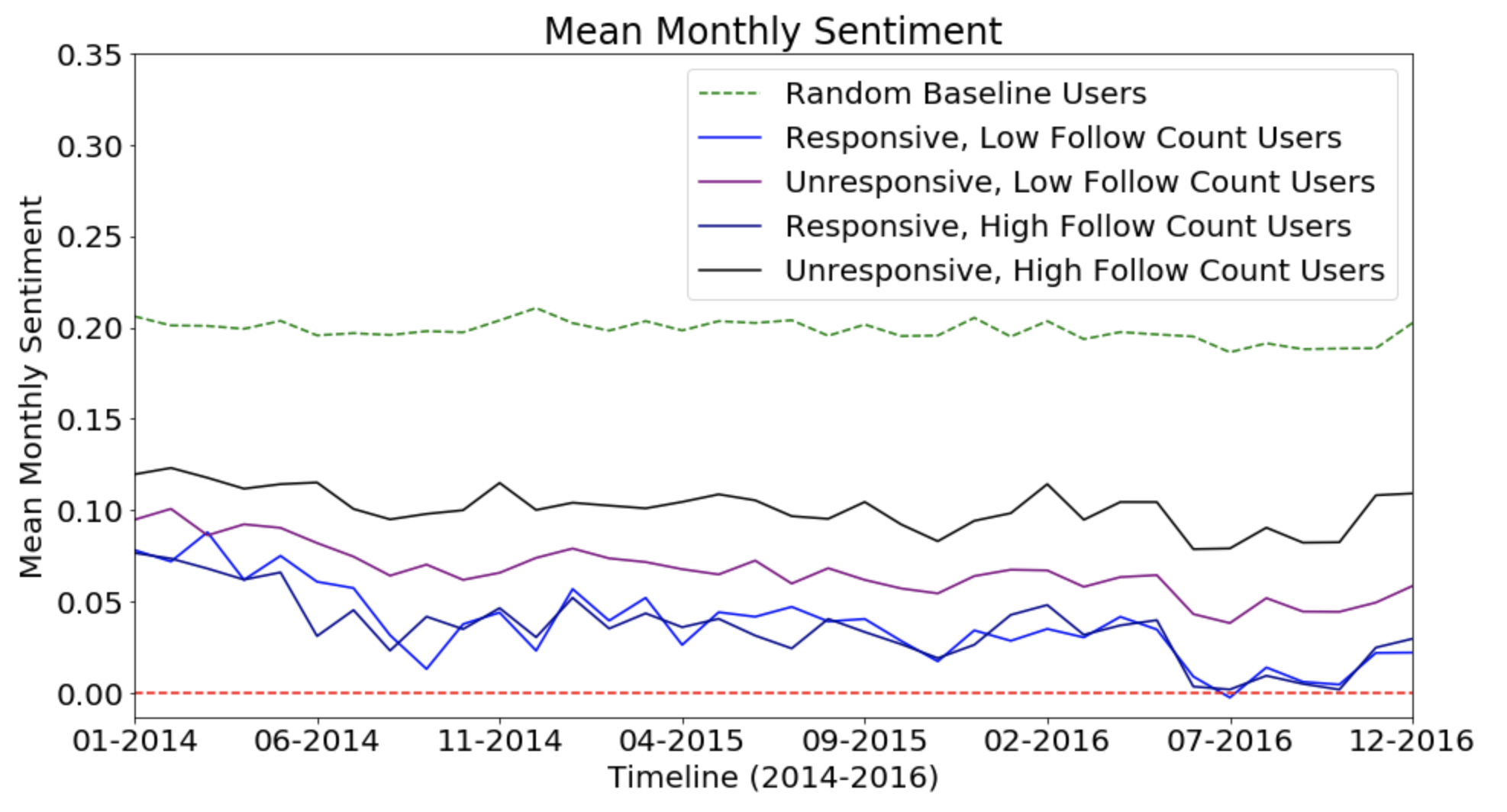}
    \caption{}
\end{subfigure}
\begin{subfigure}{.5\textwidth}
    \centering
    \includegraphics[width=1.0\textwidth]{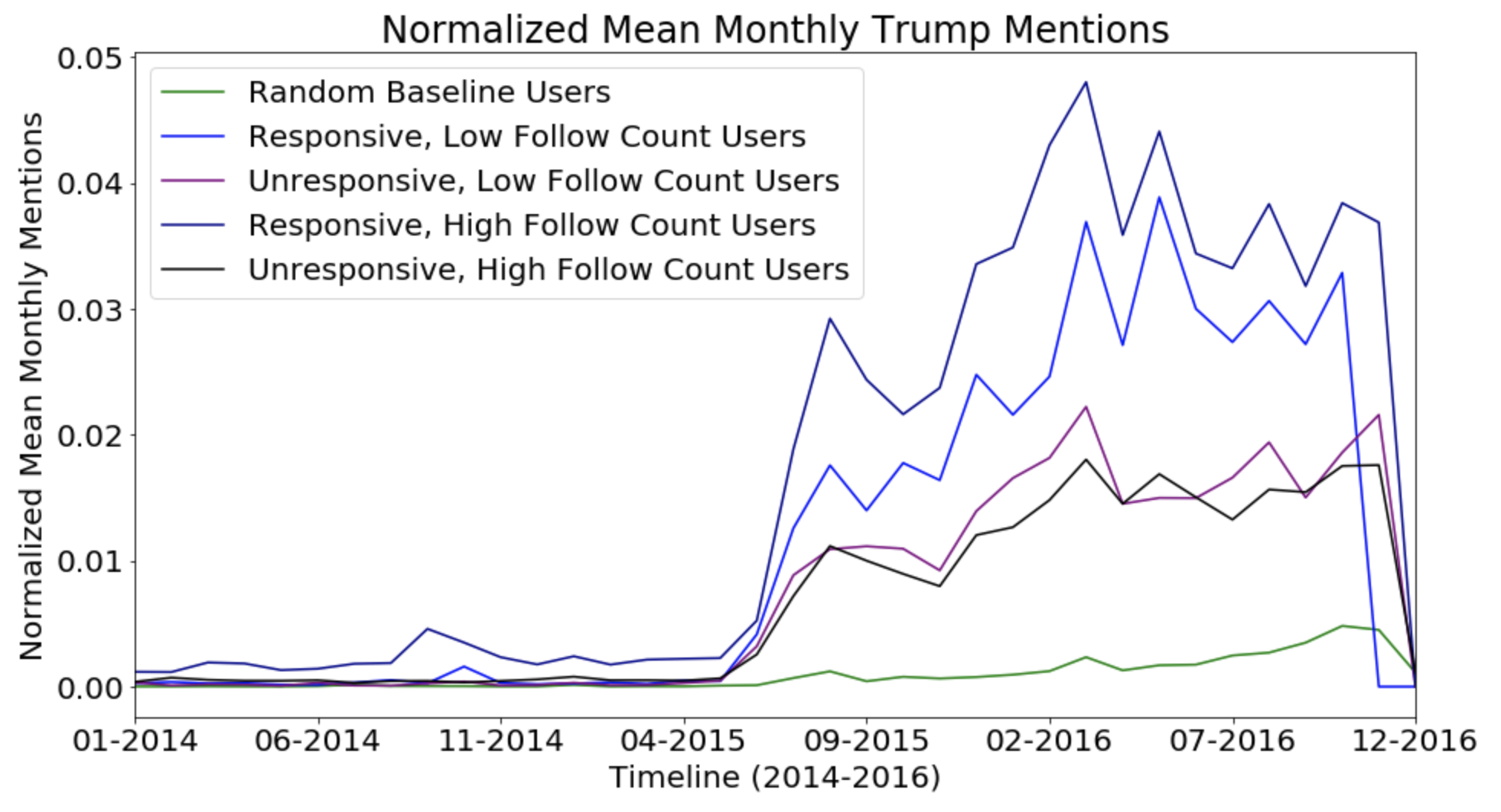}
    \caption{}
\end{subfigure}%
\begin{subfigure}{.5\textwidth}
    \centering
    \includegraphics[width=1.0\textwidth]{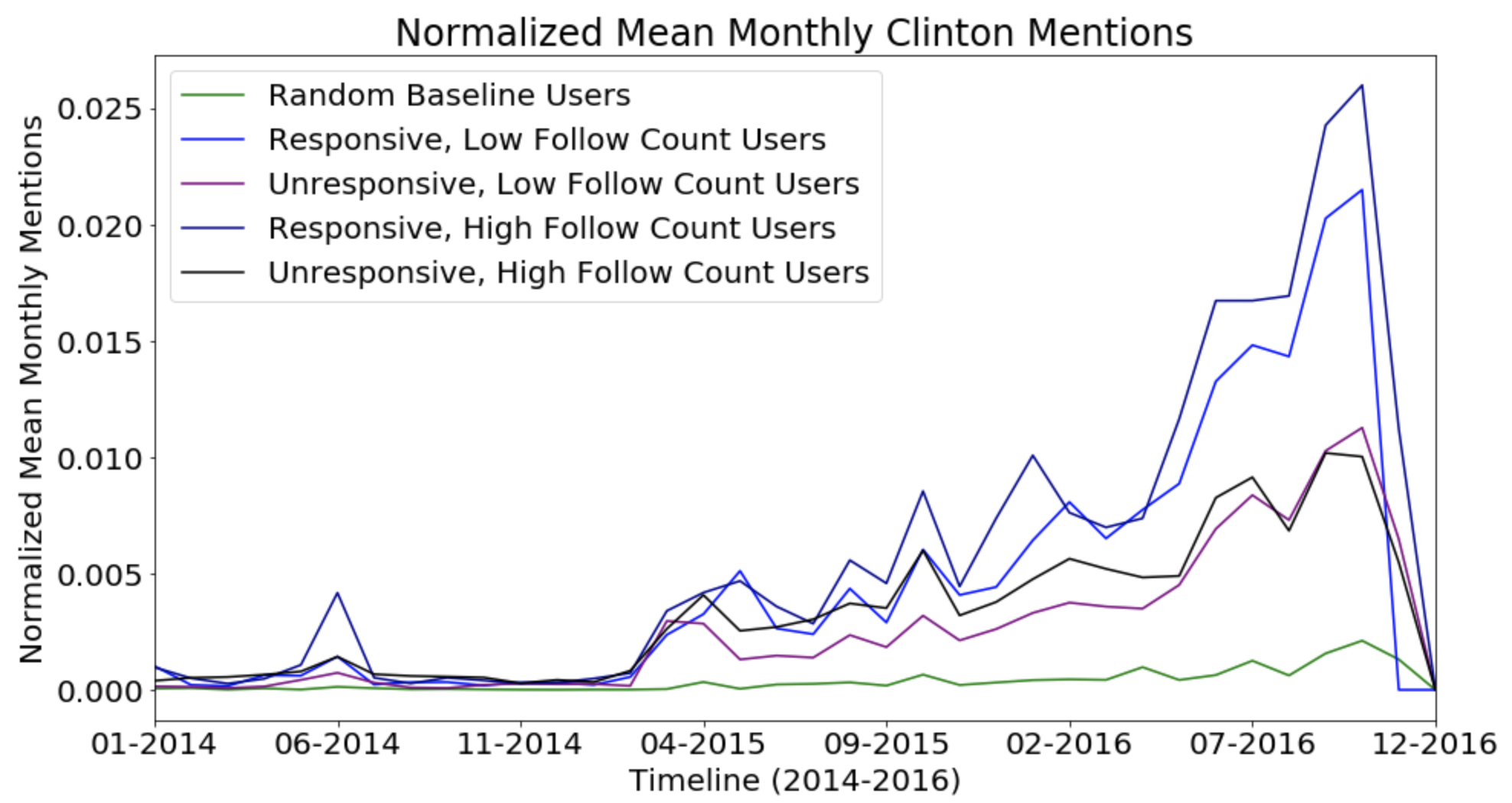}
    \caption{}
\end{subfigure}
\begin{subfigure}{.5\textwidth}
    \centering
    \includegraphics[width=1.0\textwidth]{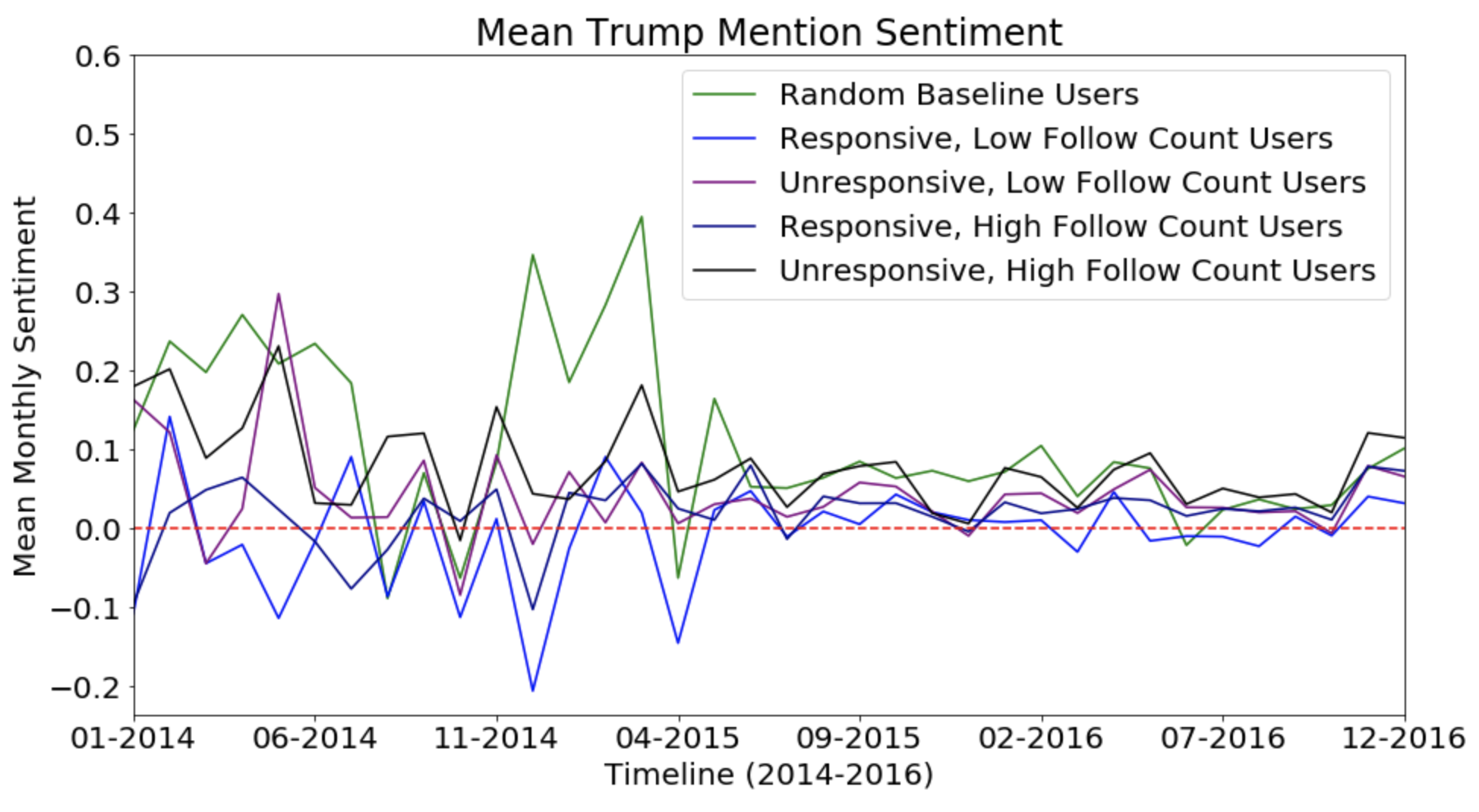}
    \caption{}
\end{subfigure}%
\begin{subfigure}{.5\textwidth}
    \centering
    \includegraphics[width=1.0\textwidth]{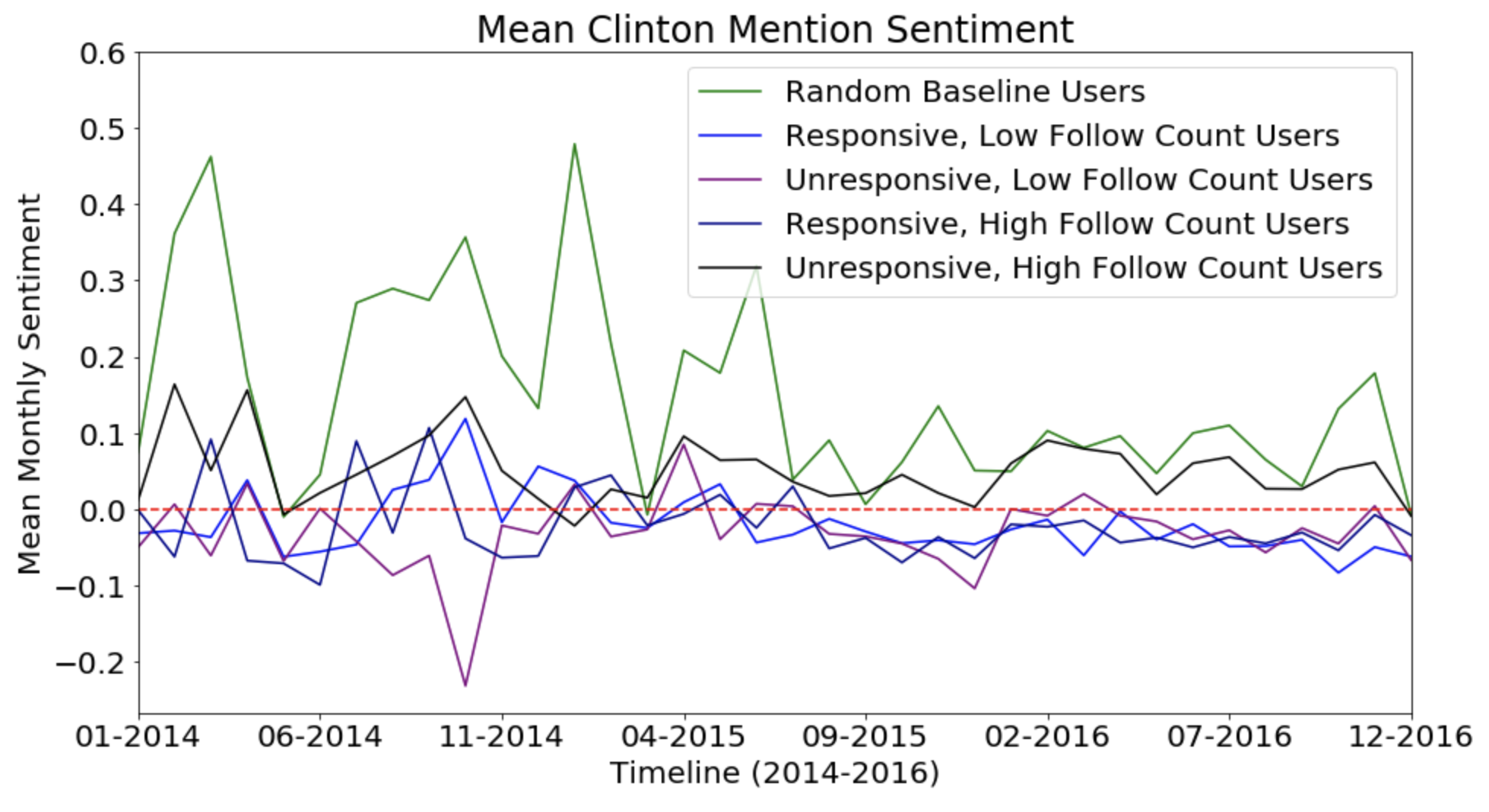}
    \caption{}
\end{subfigure}
\caption{Timeline data for the four user groups (responsive and unresponsive high follower count and low follower count users) in addition to the random baseline set of users. The depicted time span of these graphs runs from January 2014 to December 2016. The contacted users were found to be significantly different than the baseline set in all cases except for sentiment of tweets mentioning @realDonaldTrump in unresponsive high follower count users. P-values from this testing are reported in Table \ref{table2}.}
\label{timeline_fig}
\end{figure}

An analysis of behavior between the population of users which were contacted by the IRA and the random baseline over a timeline from 2009 to 2016 revealed that the contacted users significantly deviated from the random baseline in almost all behaviors. 
The Kolmogorov-Smirnov significance test was used to compare the distribution of random baseline users and that of the contacted users \cite{lilliefors1967kolmogorov}. The distribution of each group (e.g. responsive low follower count, unresponsive low follower count users, etc.) was independently tested against the baseline population. Each user group was tested against all 96 months of data collected prior to the 2016 U.S. presidential election (Jan 2009 - Dec 2016) as well as the more critical years surrounding the election (2014-2016). \textit{The KS-tests, done at 0.05 level of significance, show that all the four user groups had a significantly different distribution than that of the random baseline users aside from the sentiment of tweet mentioning @realDonaldTrump in case of the unresponsive high follow count users, when looking at behaviors from 2014 to 2016.} The p-values from this testing are reported in Table \ref{table2} and visuals are displayed in \figref{timeline_fig}. When looking at behaviors from 2009-2016, all four user groups had a significantly different distribution in all aspects than that of the random baseline users, with all p-values < 0.05.

It was found that the contacted users tweeted more frequently overall than the baseline user set \figref{timeline_fig}(a). While mentions of @realDonaldTrump and @HillaryClinton also increased within the baseline population as the US presidential election approached \figref{timeline_fig}(c,d), the contacted users mentioned these accounts more frequently still, even when counts were normalized (number of mentions divided by total tweets). The contacted users were found to be generally more negative than the baseline when considering overall sentiment \figref{timeline_fig}(b). The contacted users were also more negative in most tweets mentioning @realDonaldTrump or @HillaryClinton \figref{timeline_fig}(e,f).

\begin{table}[]
\caption{P-values returned from significance testing against the random baseline.}
\begin{tabular}{|c|c|c|}
\hline
\multicolumn{3}{|c|}{\textbf{Low Follower Count Users (2014-2016)}}                                                                                                     \\ \hline
\textbf{Behavior}                     & \textbf{Responsive} & \textbf{Unresponsive}                                      \\ \hline
Tweet Count                    & $3.73 x 10^{-17}$        & $3.73 x 10^{-17}$ \\ \hline
Overall Monthly Sentiment      & $3.73 x 10^{-17}$           & $3.73 x 10^{-17}$ \\ \hline
@realDonaldTrump Mention Count & $2.97 x 10^{-7}$              & $4.37 x 10^{-5}$ \\ \hline
@realDonaldTrump Sentiment     & $7.34 x 10^{-8}$                & 0.002 \\ \hline
@HillaryClinton Mention Count  & $7.88 x 10^{-13}$              & $3.76 x 10^{-9}$ \\ \hline
@HillaryClinton Sentiment      & $2.78 x 10^{-11}$              & $4.82 x 10^{-12}$  \\ \hline
\multicolumn{1}{l}{}                                 & \multicolumn{1}{l}{}              & \multicolumn{1}{l}{}                         \\ \hline
\multicolumn{3}{|c|}{\textbf{High Follower Count Users (2014-2016)}}                                                                                                    \\ \hline
\textbf{Behavior}                       & \textbf{Responsive} & \textbf{Unresponsive}                                     \\ \hline
Tweet Count                    & $3.73 x 10^{-17}$              & $3.73 x 10^{-17}$   \\ \hline
Overall Monthly Sentiment      & $3.73 x 10^{-17}$              & $3.73 x 10^{-17}$   \\ \hline
@realDonaldTrump Mention Count & $3.73 x 10^{-17}$             & $1.51 x 10^{-10}$    \\ \hline
@realDonaldTrump Sentiment     & $1.31 x 10^{-6}$                 & \begin{tabular}[c]{@{}c@{}}0.29\\ (NOT significant)\end{tabular}  \\ \hline
@HillaryClinton Mention Count  & $3.73 x 10^{-17}$                & $2.39 x 10^{-15}$   \\ \hline
@HillaryClinton Sentiment      & $2.78 x 10^{-11}$              & $4.82 x 10^{-12}$   \\ \hline
\end{tabular}
\label{table2}
\end{table}

\subsection{Summary of Research Findings}

\textbf{RQ1}: Did the contacted Twitter users exhibit a change in behavior following contact with IRA accounts? For all user groups, responsive and unresponsive low follower count users and high follower count users, mean tweet count significantly increased after IRA contact, as did the number of mentions of @realDonaldTrump and @HillaryClinton.  There was also a significant increase in negative sentiment for high follower count users as well as responsive low follower count users.

\textbf{RQ2}: Did the  contacted Twitter users who engaged back with the IRA accounts behave differently compared with those who did not after the contact? For certain behavioral metrics the study observed a higher percentage change in users who engaged back with the IRAs compared with those who did not, in both the low follower count and the high follower count user categories. The percentage change of tweet counts was higher in responsive users than the unresponsive ones (26.02\% and 15.43\% respectively among low follower count users and 15.96\% and 9.83\% among the high follower count users). In addition, we found that even though the low follower count responsive users as well as both the category of users in the high follower count users set show statistically significant change in overall tweet sentiments at 0.05 level of significance, at 0.01 level of significance only the low follower count users who engaged back with the IRAs show a statistically significant change, while the other user sets do not, indicating that the change in sentiments was the most prominent for the low follower count users who engaged back. 

Our findings also show that the high and low follower count users who responded back have a greater percentage change in mentioning @HillaryClinton in their tweets, compared with those who did not respond to the IRA accounts (55.44\% and 14.75\% respectively among low follower count users and 101.62\% and 67.4\% among the high follower count users). Among the low follower count users, the percentage change in mentioning @realDonaldTrump for users who responded back was greater (65.51\%) compared with the unresponsive ones (27.93\%). In contrast, among the high follower count users, the users who did not engage back showed a greater percentage change (145.82\%) compared with those who did (94.72\%).

\textbf{RQ3}: Did the behavior of users with high follower count differ from those with low follower count after initial contact with the IRA?  
We found that high and low follower count users did not exhibit sufficiently large differences in behavior according to the metrics of this study. Both high and low follower count users had similar behavioral changes in all groups for overall tweet count and mention frequency of the @realDonaldTrump and @HillaryClinton handles. All groups, except for unresponsive low follower count users, also saw a significant change in overall sentiment. No clear distinction can be made between high and low follower count users for @realDonaldTrump or @HillaryClinton sentiment.
Deeper analysis of the percentage change in each behavior reveals that while tweet count increased with greater frequency among low follower count users, high follower count users increased their uses of the @realDonaldTrump and @HillaryClinton handles more than that of the low follower count user population. For all three sentiment measures, no single set of high or low follower count users had a clearly more extreme change across both responsive and unresponsive users than the other.

%% file: conclusion.tex
\section{Discussion}

\subsection{Implications}

As it has been shown that Russia and other foreign actors have continued their efforts to sow discord around elections worldwide \cite{ghanapost, russianafrica, lawmakerswarned}, it has become more important than ever to understand the trends in behavior of users targeted in Russia's efforts. It is the hope of this work that social media platforms and governments around the globe will be spurred by the findings in this work to cooperate to implement stronger policies that protect elections from online meddling.  Given the important real-world impact of this work, our hope is also that greater collaboration between industry and academia will be stimulated.

Another potential implication of this work is that it may encourage more research investigating the effect of nefarious online political activities on the contacted user populations.  This would include studying whether users on other platforms besides Twitter, such as Facebook and Instagram, exhibited similar before/after changes in behavior.  Indeed, there is evidence that the spread of misinformation is not confined solely to Twitter \cite{facebookevidence, instaevidence, burkhardt2017combating, cooke2018fake}.  This would also include investigating which techniques achieved the greatest impact in terms of changing user behavior.  Further, while this work focused on users who were contacted by IRA accounts through mentions, retweets, and replies, i.e., the {\em intersection} of these three mechanisms, the findings may motivate further investigation of larger data sets such as the {\em union} of these three sets, which contains a much larger number of users (837,688).  Given our promising initial results, we intend to explore this wider data set in the future.  

While our work has focused on `first order' contacts between the IRA and Twitter users, this work may also raise the question of whether impacted users also impacted their friends and followers through their retweets, replies and mentions.  Thus, we encourage investigators to study the impact of `second order' contacts. 

Additional avenues of future exploration include a more in-depth analysis of user behavior. Topic modeling, for instance, would allow for a deep dive into how user-behavior on Twitter changed, beyond general attitude. It would be interesting to investigate whether these users became more or less politically inclined in the months after IRA contact, engage less with their pre-IRA contact social networks, or begin to echo the same language of IRA accounts in the month after contact, to name several possible areas of future work.





\subsection{Limitations}

We have been careful not to claim cause and effect in this research. That is, while we observe significant changes in behavior before and after contact by the IRA, we cannot definitively attribute the correlation in these changes to being primarily or even exclusively caused by contact from the IRA.  There may be other factors that caused these changes.  For example, tweet counts may have risen generally throughout the time frame considered, and sentiment may have become more negative generally and mentions of @realDonaldTrump and @HillaryClinton increased.

A limitation of this study was the inability to access several critical sources of archival Twitter data including both user retweets and favorites, on either Twint or Tweepy, which could not be accessed during the necessary time period for the majority of the users in the study \cite{twint, tweepy}. Although engaging back with the IRAs via mentions and replies can perhaps be deemed as a stronger sign of involvement with the IRAs, as compared to retweets, a log of favorite tweets as well as retweets would have allowed for a much more all-encompassing view of user behavior on the platform. 

Another limitation of this study was the inability to scientifically ascertain whether the users are Americans or not. Twitter does not allow the extraction of location information if users do not want their locations to be public. Since most of the Twitter users do not turn on their location service settings, and even fewer of them go on to actually geo-tag their tweets \cite{sloan2015tweets}, the analysis would have ended up with very few tweets having location data associated if the study had relied on tweets' location information to discern if the user is American or not. Therefore, this study focused only on users who had more than 75\% of their tweets in English.  It is unclear what fraction of these English language posters are necessarily American. The ability to access more account information or tweet metadata than Twint offers would have proven very useful in this task.




It is also difficult to ascertain user attitudes, as it does not directly correlate with sentiments, especially when working with small amounts of text, such as tweets. Although tweets mentioning @realDonaldTrump or @HillaryClinton were almost certainly pertaining to the respective candidates in some capacity, it was unclear whether the sentiment was directed at the individuals, their detractors, or some other third party.

Additionally, although the users which together form the random baseline used in this work exhibit behavior metrics which reflect those of the average Twitter user, another baseline of users not contacted by the IRA, but engaging with similar topics on Twitter would likely be the most effective user group to compare with users contacted by the IRA. It should also be noted, that the snowball sampling technique which was used to collect some three million usernames, bias could have been introduced through the four initial seeds selected as the root of sampling. In future work, topic modeling will be used to find users which have engaged with similar topics on Twitter in the hope of building a more similar group to compare against.

\section{Conclusions}

This paper has investigated the behavior of Twitter users before and after being contacted by Russia's Internet Research Agency prior to the 2016 U.S. presidential election.  Our key finding was that for all user groups of highly contacted users---namely responsive and unresponsive low follower count users and high follower count users---there were significant increases after first IRA contact in mean tweet count, and the number of mentions of @realDonaldTrump and @HillaryClinton.  There was also a significant increase in negative sentiment for high follower count users as well as responsive low follower count users.
Another key finding was that responsive users who engaged with the IRA generally showed change that was either higher in percentage or satisfied stricter statistical significance compared with non-responsive users across a variety of different metrics. Comparison with a random baseline of Twitter users has shown that contacted users had sufficiently deviated behavior from the random population in almost all areas.

%% file: acknowledgments.tex
\section{Acknowledgments}

We thank anonymous reviewers for helpful critiques and suggestions. This work is supported
in part by the US National Science Foundation (NSF) through grant CNS 1528138.

%% file: main.bbl

\begin{thebibliography}{49}


\ifx \showCODEN    \undefined \def \showCODEN     #1{\unskip}     \fi
\ifx \showDOI      \undefined \def \showDOI       #1{#1}\fi
\ifx \showISBNx    \undefined \def \showISBNx     #1{\unskip}     \fi
\ifx \showISBNxiii \undefined \def \showISBNxiii  #1{\unskip}     \fi
\ifx \showISSN     \undefined \def \showISSN      #1{\unskip}     \fi
\ifx \showLCCN     \undefined \def \showLCCN      #1{\unskip}     \fi
\ifx \shownote     \undefined \def \shownote      #1{#1}          \fi
\ifx \showarticletitle \undefined \def \showarticletitle #1{#1}   \fi
\ifx \showURL      \undefined \def \showURL       {\relax}        \fi
\providecommand\bibfield[2]{#2}
\providecommand\bibinfo[2]{#2}
\providecommand\natexlab[1]{#1}
\providecommand\showeprint[2][]{arXiv:#2}

\bibitem[\protect\citeauthoryear{Adam~Goldman and Fandos}{Adam~Goldman and
  Fandos}{2020}]%
        {lawmakerswarned}
\bibfield{author}{\bibinfo{person}{Maggie~Haberman Adam~Goldman, Julian
  E.~Barnes} {and} \bibinfo{person}{Nicholas Fandos}.}
  \bibinfo{year}{2020}\natexlab{}.
\newblock \bibinfo{booktitle}{\emph{Lawmakers Are Warned That Russia Is
  Meddling to Re-elect Trump}}.
\newblock
\urldef\tempurl%
\url{https://www.nytimes.com/2020/02/20/us/politics/russian-interference-trump-democrats.html}
\showURL{%
\tempurl}
\newblock
\shownote{[Online; accessed 15-May-2020].}


\bibitem[\protect\citeauthoryear{Albadi, Kurdi, and Mishra}{Albadi
  et~al\mbox{.}}{2019}]%
        {albadi2019hateful}
\bibfield{author}{\bibinfo{person}{Nuha Albadi}, \bibinfo{person}{Maram Kurdi},
  {and} \bibinfo{person}{Shivakant Mishra}.} \bibinfo{year}{2019}\natexlab{}.
\newblock \showarticletitle{Hateful People or Hateful Bots? Detection and
  Characterization of Bots Spreading Religious Hatred in Arabic Social Media}.
\newblock \bibinfo{journal}{\emph{Proceedings of the ACM on Human-Computer
  Interaction}} \bibinfo{volume}{3}, \bibinfo{number}{CSCW}
  (\bibinfo{year}{2019}), \bibinfo{pages}{1--25}.
\newblock


\bibitem[\protect\citeauthoryear{Arnaudo}{Arnaudo}{2017}]%
        {arnaudo2017computational}
\bibfield{author}{\bibinfo{person}{Dan Arnaudo}.}
  \bibinfo{year}{2017}\natexlab{}.
\newblock \showarticletitle{Computational propaganda in Brazil: Social bots
  during elections}.
\newblock  (\bibinfo{year}{2017}).
\newblock


\bibitem[\protect\citeauthoryear{Bail, Guay, Maloney, Combs, Hillygus, Merhout,
  Freelon, and Volfovsky}{Bail et~al\mbox{.}}{2020}]%
        {bail2020assessing}
\bibfield{author}{\bibinfo{person}{Christopher~A Bail}, \bibinfo{person}{Brian
  Guay}, \bibinfo{person}{Emily Maloney}, \bibinfo{person}{Aidan Combs},
  \bibinfo{person}{D~Sunshine Hillygus}, \bibinfo{person}{Friedolin Merhout},
  \bibinfo{person}{Deen Freelon}, {and} \bibinfo{person}{Alexander Volfovsky}.}
  \bibinfo{year}{2020}\natexlab{}.
\newblock \showarticletitle{Assessing the Russian Internet Research Agency's
  impact on the political attitudes and behaviors of American Twitter users in
  late 2017}.
\newblock \bibinfo{journal}{\emph{Proceedings of the National Academy of
  Sciences}} \bibinfo{volume}{117}, \bibinfo{number}{1} (\bibinfo{year}{2020}).
\newblock


\bibitem[\protect\citeauthoryear{Banjo}{Banjo}{2019}]%
        {washingtonpost}
\bibfield{author}{\bibinfo{person}{Shelly Banjo}.}
  \bibinfo{year}{2019}\natexlab{}.
\newblock \bibinfo{booktitle}{\emph{Facebook, Twitter and the Digital
  Disinformation Mess}}.
\newblock
\urldef\tempurl%
\url{https://www.washingtonpost.com/business/facebook-twitter-and-the-digital-disinformation-mess/2019/10/01/53334c08-e4b4-11e9-b0a6-3d03721b85ef_story.html}
\showURL{%
\tempurl}


\bibitem[\protect\citeauthoryear{Benigni, Joseph, and Carley}{Benigni
  et~al\mbox{.}}{2017}]%
        {benigni2017online}
\bibfield{author}{\bibinfo{person}{Matthew~C Benigni}, \bibinfo{person}{Kenneth
  Joseph}, {and} \bibinfo{person}{Kathleen~M Carley}.}
  \bibinfo{year}{2017}\natexlab{}.
\newblock \showarticletitle{Online extremism and the communities that sustain
  it: Detecting the ISIS supporting community on Twitter}.
\newblock \bibinfo{journal}{\emph{PloS one}} \bibinfo{volume}{12},
  \bibinfo{number}{12} (\bibinfo{year}{2017}).
\newblock


\bibitem[\protect\citeauthoryear{Bessi and Ferrara}{Bessi and Ferrara}{2016}]%
        {bessi2016social}
\bibfield{author}{\bibinfo{person}{Alessandro Bessi} {and}
  \bibinfo{person}{Emilio Ferrara}.} \bibinfo{year}{2016}\natexlab{}.
\newblock \showarticletitle{Social bots distort the 2016 US Presidential
  election online discussion}.
\newblock \bibinfo{journal}{\emph{First Monday}} \bibinfo{volume}{21},
  \bibinfo{number}{11-7} (\bibinfo{year}{2016}).
\newblock


\bibitem[\protect\citeauthoryear{Boatwright, Linvill, and Warren}{Boatwright
  et~al\mbox{.}}{2018}]%
        {boatwright2018troll}
\bibfield{author}{\bibinfo{person}{Brandon~C Boatwright},
  \bibinfo{person}{Darren~L Linvill}, {and} \bibinfo{person}{Patrick~L
  Warren}.} \bibinfo{year}{2018}\natexlab{}.
\newblock \showarticletitle{Troll factories: The internet research agency and
  state-sponsored agenda building}.
\newblock \bibinfo{journal}{\emph{Resource Centre on Media Freedom in Europe}}
  (\bibinfo{year}{2018}).
\newblock


\bibitem[\protect\citeauthoryear{Boyd, Spangher, Fourney, Nushi, Ranade,
  Pennebaker, and Horvitz}{Boyd et~al\mbox{.}}{2018}]%
        {boyd2018characterizing}
\bibfield{author}{\bibinfo{person}{Ryan~L Boyd}, \bibinfo{person}{Alexander
  Spangher}, \bibinfo{person}{Adam Fourney}, \bibinfo{person}{Besmira Nushi},
  \bibinfo{person}{Gireeja Ranade}, \bibinfo{person}{James Pennebaker}, {and}
  \bibinfo{person}{Eric Horvitz}.} \bibinfo{year}{2018}\natexlab{}.
\newblock \showarticletitle{Characterizing the Internet Research Agency's
  Social Media Operations During the 2016 US Presidential Election using
  Linguistic Analyses}.
\newblock  (\bibinfo{year}{2018}).
\newblock


\bibitem[\protect\citeauthoryear{Burkhardt}{Burkhardt}{2017}]%
        {burkhardt2017combating}
\bibfield{author}{\bibinfo{person}{Joanna~M Burkhardt}.}
  \bibinfo{year}{2017}\natexlab{}.
\newblock \bibinfo{booktitle}{\emph{Combating fake news in the digital age}}.
  Vol.~\bibinfo{volume}{53}.
\newblock \bibinfo{publisher}{American Library Association}.
\newblock


\bibitem[\protect\citeauthoryear{Clarissa~Ward and Lister}{Clarissa~Ward and
  Lister}{2020}]%
        {ghanapost}
\bibfield{author}{\bibinfo{person}{Sebastian Shukla Gianluca~Mezzofiore
  Clarissa~Ward, Katie~Polglase} {and} \bibinfo{person}{Tim Lister}.}
  \bibinfo{year}{2020}\natexlab{}.
\newblock \bibinfo{booktitle}{\emph{Russian election meddling is back -- via
  Ghana and Nigeria -- and in your feeds}}.
\newblock
\urldef\tempurl%
\url{https://www.cnn.com/2020/03/12/world/russia-ghana-troll-farms-2020-ward/index.html}
\showURL{%
\tempurl}
\newblock
\shownote{[Online; accessed 15-May-2020].}


\bibitem[\protect\citeauthoryear{Cooke}{Cooke}{2018}]%
        {cooke2018fake}
\bibfield{author}{\bibinfo{person}{Nicole~A Cooke}.}
  \bibinfo{year}{2018}\natexlab{}.
\newblock \bibinfo{booktitle}{\emph{Fake news and alternative facts:
  Information literacy in a post-truth era}}.
\newblock \bibinfo{publisher}{American Library Association}.
\newblock


\bibitem[\protect\citeauthoryear{Cresci, Di~Pietro, Petrocchi, Spognardi, and
  Tesconi}{Cresci et~al\mbox{.}}{2017}]%
        {Cresci_2017}
\bibfield{author}{\bibinfo{person}{Stefano Cresci}, \bibinfo{person}{Roberto
  Di~Pietro}, \bibinfo{person}{Marinella Petrocchi}, \bibinfo{person}{Angelo
  Spognardi}, {and} \bibinfo{person}{Maurizio Tesconi}.}
  \bibinfo{year}{2017}\natexlab{}.
\newblock \showarticletitle{The Paradigm-Shift of Social Spambots}.
\newblock \bibinfo{journal}{\emph{Proceedings of the 26th International
  Conference on World Wide Web Companion - WWW ’17 Companion}}
  (\bibinfo{year}{2017}).
\newblock
\showISBNx{9781450349147}
\urldef\tempurl%
\url{https://doi.org/10.1145/3041021.3055135}
\showDOI{\tempurl}


\bibitem[\protect\citeauthoryear{Danilak}{Danilak}{2020}]%
        {langdetect}
\bibfield{author}{\bibinfo{person}{Michal~Mimino Danilak}.}
  \bibinfo{year}{2020}\natexlab{}.
\newblock \bibinfo{booktitle}{\emph{langdetect 1.0.8}}.
\newblock
\urldef\tempurl%
\url{https://pypi.org/project/langdetect/}
\showURL{%
\tempurl}
\newblock
\shownote{[Online; accessed 15-May-2020].}


\bibitem[\protect\citeauthoryear{DiResta, Shaffer, Ruppel, Sullivan, Matney,
  Fox, Albright, and Johnson}{DiResta et~al\mbox{.}}{2018}]%
        {diresta2018tactics}
\bibfield{author}{\bibinfo{person}{Renee DiResta}, \bibinfo{person}{Kris
  Shaffer}, \bibinfo{person}{Becky Ruppel}, \bibinfo{person}{David Sullivan},
  \bibinfo{person}{Robert Matney}, \bibinfo{person}{Ryan Fox},
  \bibinfo{person}{Jonathan Albright}, {and} \bibinfo{person}{Ben Johnson}.}
  \bibinfo{year}{2018}\natexlab{}.
\newblock \showarticletitle{The tactics and tropes of the Internet Research
  Agency}.
\newblock  (\bibinfo{year}{2018}).
\newblock


\bibitem[\protect\citeauthoryear{Ferrara}{Ferrara}{2017}]%
        {ferrara2017disinformation}
\bibfield{author}{\bibinfo{person}{Emilio Ferrara}.}
  \bibinfo{year}{2017}\natexlab{}.
\newblock \showarticletitle{Disinformation and social bot operations in the run
  up to the 2017 French presidential election}.
\newblock \bibinfo{journal}{\emph{arXiv preprint arXiv:1707.00086}}
  (\bibinfo{year}{2017}).
\newblock


\bibitem[\protect\citeauthoryear{Ferrara, Varol, Davis, Menczer, and
  Flammini}{Ferrara et~al\mbox{.}}{2016}]%
        {ferrara2016rise}
\bibfield{author}{\bibinfo{person}{Emilio Ferrara}, \bibinfo{person}{Onur
  Varol}, \bibinfo{person}{Clayton Davis}, \bibinfo{person}{Filippo Menczer},
  {and} \bibinfo{person}{Alessandro Flammini}.}
  \bibinfo{year}{2016}\natexlab{}.
\newblock \showarticletitle{The rise of social bots}.
\newblock \bibinfo{journal}{\emph{Commun. ACM}} \bibinfo{volume}{59},
  \bibinfo{number}{7} (\bibinfo{year}{2016}), \bibinfo{pages}{96--104}.
\newblock


\bibitem[\protect\citeauthoryear{Gadde and Beykpour}{Gadde and
  Beykpour}{2018}]%
        {twittershadow}
\bibfield{author}{\bibinfo{person}{Vijaya Gadde} {and} \bibinfo{person}{Kayvon
  Beykpour}.} \bibinfo{year}{2018}\natexlab{}.
\newblock \bibinfo{booktitle}{\emph{Setting the record straight on shadow
  banning}}.
\newblock
\urldef\tempurl%
\url{https://blog.twitter.com/en_us/topics/company/2018/Setting-the-record-straight-on-shadow-banning.html}
\showURL{%
\tempurl}


\bibitem[\protect\citeauthoryear{Goodman}{Goodman}{1961}]%
        {10.2307/2237615}
\bibfield{author}{\bibinfo{person}{Leo~A. Goodman}.}
  \bibinfo{year}{1961}\natexlab{}.
\newblock \showarticletitle{Snowball Sampling}.
\newblock \bibinfo{journal}{\emph{The Annals of Mathematical Statistics}}
  \bibinfo{volume}{32}, \bibinfo{number}{1} (\bibinfo{year}{1961}),
  \bibinfo{pages}{148--170}.
\newblock
\showISSN{00034851}
\urldef\tempurl%
\url{http://www.jstor.org/stable/2237615}
\showURL{%
\tempurl}


\bibitem[\protect\citeauthoryear{Graff}{Graff}{2018}]%
        {wiredpost}
\bibfield{author}{\bibinfo{person}{Garrett~M. Graff}.}
  \bibinfo{year}{2018}\natexlab{}.
\newblock \bibinfo{booktitle}{\emph{Russian Trolls Are Still Playing Both
  Sides—Even With the Mueller Probe}}.
\newblock
\urldef\tempurl%
\url{https://www.wired.com/story/russia-indictment-twitter-facebook-play-both-sides/}
\showURL{%
\tempurl}


\bibitem[\protect\citeauthoryear{Howard, Ganesh, Liotsiou, Kelly, and
  Fran{\c{c}}ois}{Howard et~al\mbox{.}}{2019}]%
        {howardira}
\bibfield{author}{\bibinfo{person}{Philip~N Howard}, \bibinfo{person}{Bharath
  Ganesh}, \bibinfo{person}{Dimitra Liotsiou}, \bibinfo{person}{John Kelly},
  {and} \bibinfo{person}{Camille Fran{\c{c}}ois}.}
  \bibinfo{year}{2019}\natexlab{}.
\newblock \showarticletitle{The IRA, Social Media and Political Polarization in
  the United States, 2012-2018}.
\newblock \bibinfo{journal}{\emph{U.S. Senate Documents}}
  (\bibinfo{year}{2019}).
\newblock


\bibitem[\protect\citeauthoryear{Howard and Kollanyi}{Howard and
  Kollanyi}{2016}]%
        {howard2016bots}
\bibfield{author}{\bibinfo{person}{Philip~N Howard} {and}
  \bibinfo{person}{Bence Kollanyi}.} \bibinfo{year}{2016}\natexlab{}.
\newblock \showarticletitle{Bots,\# StrongerIn, and\# Brexit: computational
  propaganda during the UK-EU referendum}.
\newblock \bibinfo{journal}{\emph{Available at SSRN 2798311}}
  (\bibinfo{year}{2016}).
\newblock


\bibitem[\protect\citeauthoryear{Hutto and Gilbert}{Hutto and Gilbert}{2014}]%
        {vader}
\bibfield{author}{\bibinfo{person}{C.J. Hutto} {and} \bibinfo{person}{E.E.
  Gilbert}.} \bibinfo{year}{2014}\natexlab{}.
\newblock \bibinfo{booktitle}{\emph{VADER: A Parsimonious Rule-based Model for
  Sentiment Analysis of Social Media Text. Eighth International Conference on
  Weblogs and Social Media}}.
\newblock
\urldef\tempurl%
\url{https://pypi.org/project/vaderSentiment/}
\showURL{%
\tempurl}
\newblock
\shownote{[Online; accessed 15-May-2020].}


\bibitem[\protect\citeauthoryear{Keller, Schoch, Stier, and Yang}{Keller
  et~al\mbox{.}}{2020}]%
        {keller2020political}
\bibfield{author}{\bibinfo{person}{Franziska~B Keller}, \bibinfo{person}{David
  Schoch}, \bibinfo{person}{Sebastian Stier}, {and} \bibinfo{person}{JungHwan
  Yang}.} \bibinfo{year}{2020}\natexlab{}.
\newblock \showarticletitle{Political Astroturfing on Twitter: How to
  coordinate a disinformation Campaign}.
\newblock \bibinfo{journal}{\emph{Political Communication}}
  \bibinfo{volume}{37}, \bibinfo{number}{2} (\bibinfo{year}{2020}),
  \bibinfo{pages}{256--280}.
\newblock


\bibitem[\protect\citeauthoryear{Kolo and Haumer}{Kolo and Haumer}{2018}]%
        {kolo2018social}
\bibfield{author}{\bibinfo{person}{Castulus Kolo} {and}
  \bibinfo{person}{Florian Haumer}.} \bibinfo{year}{2018}\natexlab{}.
\newblock \showarticletitle{Social media celebrities as influencers in brand
  communication: An empirical study on influencer content, its advertising
  relevance and audience expectations}.
\newblock \bibinfo{journal}{\emph{Journal of Digital \& Social Media
  Marketing}} \bibinfo{volume}{6}, \bibinfo{number}{3} (\bibinfo{year}{2018}),
  \bibinfo{pages}{273--282}.
\newblock


\bibitem[\protect\citeauthoryear{Kriel and Pavliuc}{Kriel and Pavliuc}{2019}]%
        {kriel2019reverse}
\bibfield{author}{\bibinfo{person}{Charles Kriel} {and} \bibinfo{person}{Alexa
  Pavliuc}.} \bibinfo{year}{2019}\natexlab{}.
\newblock \showarticletitle{Reverse Engineering Russian Internet Research
  Agency Tactics through Network Analysis}.
\newblock \bibinfo{journal}{\emph{Defence Strategic Communication}}
  (\bibinfo{year}{2019}), \bibinfo{pages}{199--227}.
\newblock


\bibitem[\protect\citeauthoryear{Lilliefors}{Lilliefors}{1967}]%
        {lilliefors1967kolmogorov}
\bibfield{author}{\bibinfo{person}{Hubert~W Lilliefors}.}
  \bibinfo{year}{1967}\natexlab{}.
\newblock \showarticletitle{On the Kolmogorov-Smirnov test for normality with
  mean and variance unknown}.
\newblock \bibinfo{journal}{\emph{Journal of the American statistical
  Association}} \bibinfo{volume}{62}, \bibinfo{number}{318}
  (\bibinfo{year}{1967}), \bibinfo{pages}{399--402}.
\newblock


\bibitem[\protect\citeauthoryear{Lomas}{Lomas}{2020}]%
        {instaevidence}
\bibfield{author}{\bibinfo{person}{Natasha Lomas}.}
  \bibinfo{year}{2020}\natexlab{}.
\newblock \bibinfo{booktitle}{\emph{Instagram says growth hackers are behind
  spate of fake Stories}}.
\newblock
\urldef\tempurl%
\url{https://techcrunch.com/2019/08/15/instagram-says-growth-hackers-are-behind-spate-of-fake-stories-views/}
\showURL{%
\tempurl}
\newblock
\shownote{[Online; accessed 15-May-2020].}


\bibitem[\protect\citeauthoryear{Lou and Yuan}{Lou and Yuan}{2019}]%
        {lou2019influencer}
\bibfield{author}{\bibinfo{person}{Chen Lou} {and} \bibinfo{person}{Shupei
  Yuan}.} \bibinfo{year}{2019}\natexlab{}.
\newblock \showarticletitle{Influencer marketing: how message value and
  credibility affect consumer trust of branded content on social media}.
\newblock \bibinfo{journal}{\emph{Journal of Interactive Advertising}}
  \bibinfo{volume}{19}, \bibinfo{number}{1} (\bibinfo{year}{2019}),
  \bibinfo{pages}{58--73}.
\newblock


\bibitem[\protect\citeauthoryear{Mary~Ilyushina and Lister}{Mary~Ilyushina and
  Lister}{2019}]%
        {russianafrica}
\bibfield{author}{\bibinfo{person}{Gianluca~Mezzofiore Mary~Ilyushina} {and}
  \bibinfo{person}{Tim Lister}.} \bibinfo{year}{2019}\natexlab{}.
\newblock \bibinfo{booktitle}{\emph{Russia's 'troll factory' is alive and well
  in Africa}}.
\newblock
\urldef\tempurl%
\url{https://www.cnn.com/2019/10/31/europe/russia-africa-propaganda-intl/index.html}
\showURL{%
\tempurl}
\newblock
\shownote{[Online; accessed 15-May-2020].}


\bibitem[\protect\citeauthoryear{Massey~Jr}{Massey~Jr}{1951}]%
        {massey1951kolmogorov}
\bibfield{author}{\bibinfo{person}{Frank~J Massey~Jr}.}
  \bibinfo{year}{1951}\natexlab{}.
\newblock \showarticletitle{The Kolmogorov-Smirnov test for goodness of fit}.
\newblock \bibinfo{journal}{\emph{Journal of the American statistical
  Association}} \bibinfo{volume}{46}, \bibinfo{number}{253}
  (\bibinfo{year}{1951}), \bibinfo{pages}{68--78}.
\newblock


\bibitem[\protect\citeauthoryear{Mueller and Cat}{Mueller and Cat}{2019}]%
        {mueller2019report}
\bibfield{author}{\bibinfo{person}{Robert~S Mueller} {and} \bibinfo{person}{Man
  With~A. Cat}.} \bibinfo{year}{2019}\natexlab{}.
\newblock \bibinfo{booktitle}{\emph{Report on the investigation into Russian
  interference in the 2016 presidential election}}. Vol.~\bibinfo{volume}{1}.
\newblock \bibinfo{publisher}{US Department of Justice Washington, DC}.
\newblock


\bibitem[\protect\citeauthoryear{of~Representatives Permanent Select
  Committee~on Intelligence}{of~Representatives Permanent Select Committee~on
  Intelligence}{2018}]%
        {us2018exposing}
\bibfield{author}{\bibinfo{person}{US~House of~Representatives Permanent Select
  Committee~on Intelligence}.} \bibinfo{year}{2018}\natexlab{}.
\newblock \showarticletitle{Exposing Russia's Effort to Sow Discord Online: The
  Internet Research Agency and Advertisements}.
\newblock  (\bibinfo{year}{2018}).
\newblock


\bibitem[\protect\citeauthoryear{O'Sullivan}{O'Sullivan}{2020}]%
        {facebookevidence}
\bibfield{author}{\bibinfo{person}{Donie O'Sullivan}.}
  \bibinfo{year}{2020}\natexlab{}.
\newblock \bibinfo{booktitle}{\emph{Facebook: Russian trolls are back. And
  they're here to meddle with 2020}}.
\newblock
\urldef\tempurl%
\url{https://www.cnn.com/2019/10/21/tech/russia-instagram-accounts-2020-election/index.html}
\showURL{%
\tempurl}
\newblock
\shownote{[Online; accessed 15-May-2020].}


\bibitem[\protect\citeauthoryear{Policy}{Policy}{2018}]%
        {twitterpost}
\bibfield{author}{\bibinfo{person}{Twitter~Public Policy}.}
  \bibinfo{year}{2018}\natexlab{}.
\newblock \bibinfo{title}{Update on Twitter's review of the 2016 US election}.
\newblock
  \bibinfo{howpublished}{\url{https://blog.twitter.com/en_us/topics/company/2018/2016-election-update.html}}.
\newblock
\newblock
\shownote{[Online; accessed 15-May-2020].}


\bibitem[\protect\citeauthoryear{Roesslein}{Roesslein}{2019}]%
        {tweepy}
\bibfield{author}{\bibinfo{person}{Joshua Roesslein}.}
  \bibinfo{year}{2019}\natexlab{}.
\newblock \bibinfo{booktitle}{\emph{Tweepy Documentation}}.
\newblock
\urldef\tempurl%
\url{http://docs.tweepy.org/en/latest/}
\showURL{%
\tempurl}
\newblock
\shownote{[Online; accessed 15-May-2020].}


\bibitem[\protect\citeauthoryear{Ruck, Rice, Borycz, and Bentley}{Ruck
  et~al\mbox{.}}{2019}]%
        {ruck2019internet}
\bibfield{author}{\bibinfo{person}{Damian~J Ruck}, \bibinfo{person}{Natalie~M
  Rice}, \bibinfo{person}{Joshua Borycz}, {and} \bibinfo{person}{R~Alexander
  Bentley}.} \bibinfo{year}{2019}\natexlab{}.
\newblock \showarticletitle{Internet Research Agency Twitter activity predicted
  2016 US election polls}.
\newblock \bibinfo{journal}{\emph{First Monday}} \bibinfo{volume}{24},
  \bibinfo{number}{7} (\bibinfo{year}{2019}).
\newblock


\bibitem[\protect\citeauthoryear{Shao, Ciampaglia, Varol, Flammini, and
  Menczer}{Shao et~al\mbox{.}}{2017}]%
        {shao2017spread}
\bibfield{author}{\bibinfo{person}{Chengcheng Shao},
  \bibinfo{person}{Giovanni~Luca Ciampaglia}, \bibinfo{person}{Onur Varol},
  \bibinfo{person}{Alessandro Flammini}, {and} \bibinfo{person}{Filippo
  Menczer}.} \bibinfo{year}{2017}\natexlab{}.
\newblock \showarticletitle{The spread of fake news by social bots}.
\newblock \bibinfo{journal}{\emph{arXiv preprint arXiv:1707.07592}}
  \bibinfo{volume}{96} (\bibinfo{year}{2017}), \bibinfo{pages}{104}.
\newblock


\bibitem[\protect\citeauthoryear{Sloan and Morgan}{Sloan and Morgan}{2015}]%
        {sloan2015tweets}
\bibfield{author}{\bibinfo{person}{Luke Sloan} {and} \bibinfo{person}{Jeffrey
  Morgan}.} \bibinfo{year}{2015}\natexlab{}.
\newblock \showarticletitle{Who tweets with their location? Understanding the
  relationship between demographic characteristics and the use of geoservices
  and geotagging on Twitter}.
\newblock \bibinfo{journal}{\emph{PloS one}} \bibinfo{volume}{10},
  \bibinfo{number}{11} (\bibinfo{year}{2015}).
\newblock


\bibitem[\protect\citeauthoryear{Solutions}{Solutions}{2020}]%
        {wilcoxon}
\bibfield{author}{\bibinfo{person}{Statistics Solutions}.}
  \bibinfo{year}{2020}\natexlab{}.
\newblock \bibinfo{booktitle}{\emph{How to Conduct the Wilcoxon Sign Test}}.
\newblock
\urldef\tempurl%
\url{https://www.statisticssolutions.com/how-to-conduct-the-wilcox-sign-test/}
\showURL{%
\tempurl}
\newblock
\shownote{[Online; accessed 15-May-2020].}


\bibitem[\protect\citeauthoryear{Stack}{Stack}{2018}]%
        {nytshadow}
\bibfield{author}{\bibinfo{person}{Liam Stack}.}
  \bibinfo{year}{2018}\natexlab{}.
\newblock \bibinfo{booktitle}{\emph{What Is a 'Shadow Ban,' and Is Twitter
  Doing It to Republican Accounts?}}
\newblock
\urldef\tempurl%
\url{https://www.nytimes.com/2018/07/26/us/politics/twitter-shadowbanning.html}
\showURL{%
\tempurl}


\bibitem[\protect\citeauthoryear{Starbird, Arif, and Wilson}{Starbird
  et~al\mbox{.}}{2019}]%
        {starbird2019disinformation}
\bibfield{author}{\bibinfo{person}{Kate Starbird}, \bibinfo{person}{Ahmer
  Arif}, {and} \bibinfo{person}{Tom Wilson}.} \bibinfo{year}{2019}\natexlab{}.
\newblock \showarticletitle{Disinformation as collaborative work: Surfacing the
  participatory nature of strategic information operations}.
\newblock \bibinfo{journal}{\emph{Proceedings of the ACM on Human-Computer
  Interaction}} \bibinfo{volume}{3}, \bibinfo{number}{CSCW}
  (\bibinfo{year}{2019}), \bibinfo{pages}{1--26}.
\newblock


\bibitem[\protect\citeauthoryear{team}{team}{2020}]%
        {twint}
\bibfield{author}{\bibinfo{person}{OSINT team}.}
  \bibinfo{year}{2020}\natexlab{}.
\newblock \bibinfo{booktitle}{\emph{TWINT - Twitter Intelligence Tool}}.
\newblock
\urldef\tempurl%
\url{https://github.com/twintproject/twint}
\showURL{%
\tempurl}
\newblock
\shownote{[Online; accessed 15-May-2020].}


\bibitem[\protect\citeauthoryear{Thompson}{Thompson}{2018}]%
        {engagementpost}
\bibfield{author}{\bibinfo{person}{Nicholas Thompson}.}
  \bibinfo{year}{2018}\natexlab{}.
\newblock \bibinfo{booktitle}{\emph{How Russian Trolls Used Meme Warfare to
  Divide America}}.
\newblock
\urldef\tempurl%
\url{https://www.wired.com/story/russia-ira-propaganda-senate-report/}
\showURL{%
\tempurl}
\newblock
\shownote{[Online; accessed 15-May-2020].}


\bibitem[\protect\citeauthoryear{Wilson and Starbird}{Wilson and
  Starbird}{2020}]%
        {wilson2020cross}
\bibfield{author}{\bibinfo{person}{Tom Wilson} {and} \bibinfo{person}{Kate
  Starbird}.} \bibinfo{year}{2020}\natexlab{}.
\newblock \showarticletitle{Cross-platform disinformation campaigns: lessons
  learned and next steps}.
\newblock \bibinfo{journal}{\emph{Harvard Kennedy School Misinformation
  Review}} \bibinfo{volume}{1}, \bibinfo{number}{1} (\bibinfo{year}{2020}).
\newblock


\bibitem[\protect\citeauthoryear{Wojcik, Messing, Smith, Rainie, and
  Hitlin}{Wojcik et~al\mbox{.}}{2018}]%
        {wojcik2018bots}
\bibfield{author}{\bibinfo{person}{Stefan Wojcik}, \bibinfo{person}{Solomon
  Messing}, \bibinfo{person}{Aaron Smith}, \bibinfo{person}{Lee Rainie}, {and}
  \bibinfo{person}{Paul Hitlin}.} \bibinfo{year}{2018}\natexlab{}.
\newblock \showarticletitle{Bots in the Twittersphere}.
\newblock \bibinfo{journal}{\emph{Pew Research Center}}  \bibinfo{volume}{18}
  (\bibinfo{year}{2018}).
\newblock


\bibitem[\protect\citeauthoryear{Woolley and Howard}{Woolley and
  Howard}{2017}]%
        {woolley2017computational}
\bibfield{author}{\bibinfo{person}{Samuel~C Woolley} {and}
  \bibinfo{person}{Philip Howard}.} \bibinfo{year}{2017}\natexlab{}.
\newblock \showarticletitle{Computational propaganda worldwide: Executive
  summary}.
\newblock  (\bibinfo{year}{2017}).
\newblock


\bibitem[\protect\citeauthoryear{Yang}{Yang}{2020}]%
        {botometer}
\bibfield{author}{\bibinfo{person}{Kaicheng Yang}.}
  \bibinfo{year}{2020}\natexlab{}.
\newblock \bibinfo{booktitle}{\emph{Botometer Python API}}.
\newblock
\urldef\tempurl%
\url{https://github.com/yangkcatiu/botometer-python}
\showURL{%
\tempurl}


\bibitem[\protect\citeauthoryear{Yang, Varol, Davis, Ferrara, Flammini, and
  Menczer}{Yang et~al\mbox{.}}{2019}]%
        {doi:10.1002/hbe2.115}
\bibfield{author}{\bibinfo{person}{Kai-Cheng Yang}, \bibinfo{person}{Onur
  Varol}, \bibinfo{person}{Clayton~A. Davis}, \bibinfo{person}{Emilio Ferrara},
  \bibinfo{person}{Alessandro Flammini}, {and} \bibinfo{person}{Filippo
  Menczer}.} \bibinfo{year}{2019}\natexlab{}.
\newblock \showarticletitle{Arming the public with artificial intelligence to
  counter social bots}.
\newblock \bibinfo{journal}{\emph{Human Behavior and Emerging Technologies}}
  \bibinfo{volume}{1}, \bibinfo{number}{1} (\bibinfo{year}{2019}),
  \bibinfo{pages}{48--61}.
\newblock
\urldef\tempurl%
\url{https://doi.org/10.1002/hbe2.115}
\showDOI{\tempurl}
\showeprint{https://onlinelibrary.wiley.com/doi/pdf/10.1002/hbe2.115}


\end{thebibliography}
